\documentclass[11pt]{article}

\usepackage{fullpage,amssymb,amsmath,amsthm,mathtools,natbib,soul,natbib,url,enumitem,calc}
\usepackage{array}
\usepackage[margin=2.5cm]{geometry}
\newcommand\x{\checkmark}
\newcommand\y{\times}
\usepackage[utf8]{inputenc}

\usepackage{tikz}
\usetikzlibrary{matrix,fit}
\usetikzlibrary{backgrounds}

\newtheorem{lemma}{Lemma}
\newtheorem{proposition}{Proposition}
\newtheorem{corollary}{Corollary}

\def\pr{{^\prime }}
\def\tall{\textsc{tall}}
\def\wide{\textsc{wide}}

\newcommand{\tF}{\tilde F}
\newcommand{\tL}{\tilde \Lambda}

\newcommand{\hF}{\tilde F^+}
\newcommand{\hFp}{{\tilde F}^{+\pr}}

\newcommand{\hL}{\tilde\Lambda^+}

\newcommand{\calE}{{\cal E}}

\newcommand{\Lambdaop}{\Lambda^{0\prime}}
\newcommand{\Fop}{ F^{0\prime}}

\def\pconv{\smash{\mathop{\longrightarrow}\limits^p}}     
\def\dconv{\smash{\mathop{\longrightarrow}\limits^d}}
\def\To{T_o}  
\def\Tm{T_m}    
\def\No{N_o}
\def\Nm{N_m}

\def\plim{\mbox{plim }}

\def\diag{\mbox{diag}}
\def\tilde{\widetilde}
\setstcolor{red}

\let\LARGE=\large
\allowdisplaybreaks

\begin{document}
\title{{\normalsize{\bf{MATRIX COMPLETION, COUNTERFACTUALS, AND FACTOR ANALYSIS OF MISSING DATA}}}}

\author{Jushan Bai\thanks{Columbia University, 420 W. 118 St. MC 3308, New York, NY 10027.
Email: jb3064@columbia.edu} \and Serena Ng\thanks{Columbia University and NBER, 420 W. 118 St. MC 3308, New York, NY 10027. Email: serena.ng@columbia.edu \newline  We thank Ercument Cahan for helpful discussions. We also thank three anonymous referees, Markous Pelger, Ruoxuan Xiong,  and participants at the Big Data and Machine Learning Conference at the University of Chicago, and the Conference on Synthetic Controls and Related Methods for helpful comments.  Yankang (Bennie) Chen provided excellent research assistance. This work is supported by the National Science Foundation SES-1658770(Bai), and SES-1558623,  2018369  (Ng).}}
\bibliographystyle{harvard}
\date{\today\bigskip}
\maketitle
\begin{abstract}
This paper proposes an imputation  procedure that uses the factors estimated  from a   \textsc{tall} block along with  the re-rotated loadings estimated from a \textsc{wide} block to impute  missing values in a panel of data.  Assuming  that  a strong  factor structure holds for the full panel of data and its sub-blocks,  it is shown that  the common component can be consistently estimated at  four different rates of convergence   without requiring regularization or  iteration.  An asymptotic analysis of the estimation error  is obtained.  An application of our analysis is  estimation of counterfactuals when  potential outcomes have a factor structure.  We  study the  estimation of  average and individual treatment effects on the treated and establish a normal distribution theory that can be useful for hypothesis testing.
\end{abstract}

\bigskip


\bigskip


\thispagestyle{empty}
\setcounter{page}{0}
\baselineskip=18.0pt

\newpage

\section{Introduction}
Missing observations are prevalent in  empirical work, and  it is not surprising that solutions have been proposed by researchers in many disciplines. The classic econometric solution is some variant  of the  EM algorithm. In the case of factor analysis with missing data, the EM approach  is to predict  the missing values using initial estimates of the  factors obtained from  a balanced panel and iterate.  While convergence of the algorithm can be established, the asymptotic properties of the converged estimates are not well understood.

Progress can be made if the panel of incompletely observed data $X$ is large in both   dimensions and  have a  strong factor structure. This means in particular that $X$ has a common component $C$ of reduced rank $r$, and whose population covariance matrix  has $r$ eigenvalues that increase with the size of the panel.  We show in this paper that in spite of missing values in $X$, every entry of  $C$ can be consistently estimated using  a   \textsc{tall-wide} (\textsc{tw}) algorithm that involves  two applications of principal components.    We  provide an asymptotic  characterization of the estimation error for each  $C_{it}$ and  show that there will be four convergence rates  depending on observability of  $X_{it}$.  The approach can be used to construct missing values of potential outcomes satisfying a factor structure.

Our factor imputation approach  contrasts with those used in  matrix completions. In that literature, the probabilistic structure of the data is not specified;  nuclear norm regularization via singular-value thresholding is crucial, and successful matrix recovery typically requires that the low rank component  is  {\em incoherent}, and that the  data are missing uniformly at random. See, for example,  \citet{cai-candes-shen}. Instead, we impose moment conditions  to ensure that the assumed factor structure is strong and identifiable.  We also require that $N_o\rightarrow \infty$ and $T_o\rightarrow\infty$, where  $N_{o}$ is the number of units with  data observed over the entire span, and $T_{o}$ is the length of the sample that data are available for all $N$ units. Exploiting the commonality between the observed and missing data allows recovery of the entire  matrix $C$  using  principal components  with as few as $N_{o}T+T_{o} N$ observations.     Provided that there are enough observed data to estimate the factors and the loadings consistently, a large  fraction  $1- \frac{N_{o}}{N}-\frac{T_{o}}{T}$ of the data can potentially be missing. And while regularization can yield fewer factors, it is not needed for consistent estimation of the missing values. In independent work, \citet{su-missing} and \citet{xiong-pelger:19}  suggest  alternative factor-based approaches. What makes our theory distinct is  that we analyze  factors estimated directly from  the data  without preliminary adjustment or iteration. The  estimates that emerge from the \textsc{tw} algorithm are already consistent and asymptotically normal, though one re-estimation using imputed data will provide a faster convergence rate for one sub-block.

An immediate application of the \textsc{tw} approach is program evaluation in which the object of interest is the  effect of treatment in the potential outcome model.      The method of synthetic control  pioneered in \citet{abadie:03}   estimates potential outcomes from  a weighted average of the control units, assuming that the difference between the treated and the control group is constant in the absence of treatment.    Common  variations between the treated and the control groups are crucial in  estimation of counterfactual outcomes,  and the factor model is a natural framework for modeling them.   Treating potential outcomes as missing values, we develop a factor-based estimator of the individual as well as the average treatment effect.  A distribution theory that permits tests of hypothesis is developed assuming that   the size of the control group is large.

 The rest of the paper is structured as follows. After presentation of the preliminaries in Section 2,  Section 3 presents the least squares version of Algorithm \textsc{tw}  and studies the asymptotic properties of the factor estimates that the algorithm delivers. Section 4 studies a factor-based estimation of  treatment effect.  Section 5 concludes.

\section{Preliminaries} \label{sec:section2}

We use  $i=1,\ldots N$ to index cross-section units and $t=1,\ldots T$ to index time series observations. Let  $X_i=(X_{i1},\ldots X_{iT})^\prime$ be a $T\times 1$ vector of random variables and
 $X=(X_1,X_2,\ldots, X_N)$ be a $T\times N$  matrix.  In practice, $X_{i}$ is transformed to be stationary, demeaned, and often standardized. The normalized data $Z=\frac{X}{\sqrt{NT}}$
has  singular value decomposition (\textsc{svd})
\[ Z=\frac{X}{\sqrt{NT}}
 =U DV\pr=U_r D_r V_r\pr+ U_{N-r}  D_{N-r} V_{N-r}\pr.\]
In the above, $ D_r$ is a diagonal matrix of $r$ singular values, $U_r, V_r$ are the corresponding left and right singular vectors respectively. Without loss of generality, the singular values in the diagonal entries of $ D_r$ are ordered such that $ d_1\ge d_2\ldots \ge d_r$. Note that while the $r$ largest singular values of $X$  diverge and the remaining $N-r$ ones are bounded, the $r$ largest singular values
 of $Z$ are bounded and the remaining ones tend to zero because the singular values of $Z$ are those of $X$ divided by $\sqrt{NT}$. The   \citet{eckart-young} theorem    posits that  the best  rank $k$ approximation of $Z$  is $U_kD_kV_k\pr$. The \textsc{svd} is  also the goto algorithm for solving  {\em matrix factorization} problems that seek to represent  a matrix $Z$  as a product of two low rank matrices. These results can be obtained without an assumed data generating process for $Z$.

 We are interested in the principal components of $X$ viewed from the perspective of a factor model. Let   $F$ be a $T\times r$ matrix of common factors,  $\Lambda$ be a $N \times r$ matrix of factor loadings, and $e$ be a $T\times N$ matrix of  idiosyncratic errors $e$.   The  data $X$ are assumed to have a factor structure
    \begin{eqnarray}
    \label{eq:dgp}
    X&=&F^0\Lambda^{0^\prime} + e
    \end{eqnarray}
 where $(F^0,\Lambda^0)$ are the true values of $(F,\Lambda)$. The common component $C^0=F^0\Lambda^{0\pr}$ has reduced rank $r$ because $F^0$ and $\Lambda^0$ both have rank $r$. The econometrics literature on large dimensional factor analysis  has largely adopted as framework the  {\em approximate factor model}  due to \citet{chamberlain-rothschild} according to which   the largest $r$ population eigenvalues of $X$ will increase with $N$ and $T$ while the remaining ones are bounded. The following assumptions  formalize these ideas in a statistical setting:
\paragraph{Assumption A:} There exists a constant $M<\infty$ not depending on $N,T$ such that
\begin{itemize}
    \item[(a)] (Factors and Loadings):
(i) $E\|F_t^0\|^4 \le M$, $\|\Lambda_i^0\|\le M$,
 (ii)
$\frac{F^{0\pr}F^0}{T}\pconv \Sigma_F >0$, and $ \frac{\Lambda^{0\pr}\Lambda^0 }{N}\pconv \Sigma_{\Lambda}>0$, and (iii) the eigenvalues of
$\Sigma_F \Sigma_{\Lambda}$ are distinct.
\item[(b)] (Idiosyncratic Errors):
\begin{itemize}
\item[(i)] $E(e_{it})=0, E|e_{it}|^8\le M$;
\item[(ii)] $E(\frac{1}{N}\sum_{i=1}^N e_{it} e_{is})=\gamma_N(s,t)$, $\sum_{t=1}^T |\gamma_N(s,t)|\le M$,  $\forall s$; 
\item[(iii)] $E(e_{it}e_{jt})=\tau_{ij,t}$, $|\tau_{ij,t}|\le | \tau_{ij}|$ for some $ \tau_{ij}$  $\forall t$, and
$\sum_{j=1}^N |\tau_{ij}|\le M$, $\forall i$;
\item[(iv)] $E(e_{it}e_{js})=\tau_{ij,st}$ and $\frac{1}{NT}\sum_{i=1}^N \sum_{j=1}^N \sum_{t=1}^T \sum_{s=1}^T|\tau_{ij,ts}|<M$;
\item[(v)]  $E|N^{-1/2}\sum_{i=1}^N  [e_{is}e_{it}-E(e_{is}e_{it})]^4\le M$ for every $(t,s)$;
\item[(vi)] $E(\frac{1}{N}\sum_{i=1}^N \|\frac{1}{\sqrt{T}} \sum_{t=1}^T F_t^0e_{it}\|^2)\le M$.
\end{itemize}
\item[(c)] (Central Limit Theorems): for each $i$ and $t$,
$   \frac{1}{\sqrt{N}} \sum_{i=1}^N \Lambda^0 _i e_{it}\dconv N(0,\Gamma_t)$ as  $N\rightarrow\infty$, and
$   \frac{1}{\sqrt{T}}\sum_{t=1}^TF^0_te_{it}\dconv N(0,\Phi_i) $ as  $T\rightarrow \infty$.
\end{itemize}

Assumption A  ensures  that the factor structure is strong and can be separated from the idiosyncratic errors.
The requirements  that   $E\|F_t^0\|^4\le M$ and $\|\Lambda_i^0\|\le M$  ensure that the eigenvectors of the low rank (common) component are sufficiently spread out and play the role analogous to  incoherence conditions.   Positive definiteness of $\Sigma_F$ and $\Sigma_\Lambda$   ensure that each factor has a non-trivial contribution to the low rank component. It is possible for A(a.i) to hold but not A(a.ii) and vice versa, and identification will fail. The assumption of distinct eigenvalues is used to separately identify the factors and factor loadings but is not needed to identify the common components.
Part (b) of Assumption A allows the errors to be weakly correlated both in the time and cross-section dimensions. For example, b(ii) requires the sum of autocovariances be bounded, and b(iii) is an analogous assumption for cross-sectional weak dependence.
Assumption A(vi) assumes weak dependence between the factors and the errors.
   Part (c) is needed for asymptotic distribution of the factor estimates.

  We observe $X$ but not $F^0$ or $\Lambda^0$, and $F$ and $\Lambda$ are not separately identifiable. We use as in \citet{stock-watson-jasa:02,baing-ecta:02,bai-ecta-03}  the normalizations  $\frac{F\pr F}{T}=I_r$ and  $\Lambda'\Lambda$  diagonal. The method of asymptotic principal components (APC) then constructs the factor estimates as
 \[ (\tilde F,\tilde \Lambda)=(\sqrt{T} U_r,\sqrt{N}V_rD_r). \]
  For each $t\in [1,T]$ and for each $i\in[1,N]$,    $(\tilde F_t$, $\tilde \Lambda_i)$ consistently estimate $(F_t^0$, $\Lambda^0_i$)   up to  rotation matrices $H$ and $G$ where
  \begin{eqnarray*}    H&=&\bigg(\frac{\Lambda^{0^\prime}\Lambda^0}{N}\bigg)\bigg(\frac{F^{0^\prime}\tilde F}{T}\bigg) D_r^{-2}, \quad \text{with} \quad G=H^{-1}.
    \end{eqnarray*}
   For generic positive integers $\mathbb N, \mathbb T$, it will be convenient to define
\begin{subequations}
   \begin{eqnarray}
     \delta_{\mathbb N\mathbb T}&=&\min(\sqrt{\mathbb N},\sqrt{\mathbb T}) \label{eq:delta}\\
     \mathbb V_{it}(\mathbb N,\mathbb T)&=& \frac{\delta^2_{\mathbb N\mathbb T}}{\mathbb N}\Lambda_i^{0\prime} \Sigma_{  \Lambda}^{-1}  \Gamma_t \Sigma_{\Lambda}^{-1}\Lambda^0_i+
     \frac{\delta^2_{\mathbb N\mathbb T}}{\mathbb T}F_t^{0\prime} \Sigma_{F}^{-1}  \Phi_i \Sigma_{F}^{-1}F^0_t\label{eq:eqVW} \\ &\equiv& \frac{\delta^2_{\mathbb N\mathbb T}}{\mathbb N} V_{it}+\frac{\delta_{\mathbb N\mathbb T}^2}{\mathbb T} W_{it}.\nonumber
   \end{eqnarray}
\end{subequations}
    \begin{lemma}
      \label{lem:bai-03}
 (from \citet{bai-ecta-03}): Suppose that Assumption A hold. Let  $\mathbb Q_r=\mathbb D_{r} \mathbb V_{r}\Sigma_{\Lambda}^{-1/2}$ where $\mathbb D_{r}^2$ is a diagonal matrix consisting of the eigenvalues of the $r\times r$ matrix $\Sigma_{\Lambda}^{1/2}\Sigma_F\Sigma_{\Lambda}^{1/2}$, and $\mathbb V_r$ is the corresponding matrix of eigenvectors. Then  $\plim_{N,T\rightarrow\infty} \frac{\tilde F\pr F^0}{T}=\mathbb Q_r,$
$\plim_{N,T\rightarrow\infty} D_r^2 = \mathbb D_r^2$.   Assume $\sqrt{N}/T\rightarrow 0$ and $\sqrt{T}/N\rightarrow 0$
as $ N,T\rightarrow \infty$, then
\begin{subequations}
\begin{eqnarray}
\sqrt{N} (\tilde F_t-  H\pr  F_t^0)&
\dconv&  \mathcal N\bigg(0,   \mathbb D_{r}^{-2}\mathbb Q_r \Gamma_t \mathbb Q_r\pr \mathbb D_{r}^{-2}\bigg)
\label{eq:F-full}\\
\sqrt{T} (\tilde \Lambda_i-  G \Lambda_i^0)&
\dconv & \mathcal N\bigg(0,  (\mathbb Q_r\pr)^{-1} \Phi_i \mathbb Q_r^{-1} \bigg)
\label{eq:Lambda-full}\\
  \delta_{NT}\bigg(\frac{\tilde C_{it}-C_{it}^0 }{\sqrt{\tilde{\mathbb V}_{it}(N,T)}}\bigg) &\dconv& \mathcal N(0,1) \label{eq:C-full}
\end{eqnarray}
where  $\tilde{\mathbb V}_{it}(N,T)$ is a consistent estimate of  $\mathbb V_{it}( N, T) $ defined in (\ref{eq:eqVW}).
\end{subequations}
\end{lemma}

When Assumption A is satisfied, the factors and loadings are consistently estimable up to a rotation, and the low rank component is recoverable. Furthermore, the number of factors can be consistently estimated using, for example, the criteria developed in  \citet{baing-ecta:02,baing-joe:19}.  Hence the number of factors $r$ can be treated as known.

\section{Missing Data}

Missing data is a problem that researchers frequently encounter. As \citet{zhu-wang-samworth:19} points out, we can expect more occurrence of incomplete observations  in the era of big data.  Data can be  missing for a variety of reasons:  non-response in surveys,  lack of economic activity,  and staggered releases by statistical agencies to name a few.   One can always work with a balanced panel but this effectively throws away information in many series  and  cannot be efficient.  This has led to development of simple methods that replace the missing values with zero or the mean as well as  sophisticated methods that fully specify the data generating process and   the missing data mechanism. For example,  \citet{rubin:87} suggests a Bayesian approach that fills in missing values by repeatedly sampling from the predictive distribution of the missing values.  \citet{kamakura-wedel} suggests a simulation based approach that is  aimed at handling different types of missing data in survey responses within a likelihood setup. See \citet{horton-kieinman} for  a survey of the literature.  \citet{rubin:76} obtains two sufficient conditions for  unbiased estimation. First,  missingness cannot depend on the missing values after conditioning on the observed data (a condition known as missing at random), and second, the parameters of the model must not depend on the missingness mechanism.   While missing at random can be a reasonable characterization in  observational studies and  surveys, there are situations when the assumption is not appropriate.  For example, high income survey respondents may be  more likely to ignore questions  with tax consequences.  

 The EM algorithm of \citet{dempster-laird-rubin:77} imputes missing  values by  alternating between an \textsc{E}-step that computes the expected log-likelihood using the most recent  parameter estimates, and  an \textsc{M}-step that maximizes the expected log-likelihood. In cases when the expected log-likelihood is difficult to compute,  the Expectation-Conditional Maximization  algorithm of \citet{meng-rubin:93} can be considered. \citet{schneider:01} considers a ridge-regression based regularized EM algorithm for imputing missing values in climate data.  But as \citet{honaker-king:10} noted, methods that work well in a cross-section setting  tend not to work well in panel data that exhibit dependence across units and over time. Imputing the missing values using the Kalman filter  such as discussed in  \citet{shumway-stoffer:82} remains a popular time series approach, but it is fully parametric. Dropping a series altogether because of partially missing data could lead to  a huge loss of information.  In the FRED-MD  database of over 130 series  for example, as many as 30 series can be discarded even though some are missing  for only  a handful of months (about 2\% of the observations in the panel) over the sample  1960:01 to 2018:12.

For estimation of strict factor models with missing data, more options are available. \citet{banbura-modugno:14}, \citet{jungbacker-koopman}, \citet{jkv:11} consider likelihood estimation which is conceptually appealing but  non-linear filters are needed to compute the likelihood as $F_t$ and $\Lambda_i$ are both random. For approximate factor models,  \citet{stock-watson-di-wp} suggests to  fill missing values in $X$  with the  most recent estimate of the common component. Though the precise implementation may differ, using  estimates   from  the balanced panel as initial values is the most common, see  \citet{stock-watson:handbook-16} and
 \citet{giannone-reichlin-small}.

While a variety of methods  have been proposed for  factor analysis with missing data, there are surprisingly few theoretical results until recently.   \citet{su-missing} puts zeros to observations assumed to be  missing at random and rescales the asymptotic principal components  by the probability of missing data. It is shown that these estimates are consistent but  not asymptotically normal in general, though iteration can restore normality.  \citet{xiong-pelger:19}  estimates the factors from a weighted covariance matrix. The estimates are robust to the unknown missing pattern  at the expense of larger  variances. Our  objective is similar to that of \citet{su-missing} and \citet{xiong-pelger:19} and also use a factor model for imputation, but we re-organize the data instead of re-weigh them, and we  do not make assumptions about the missing data mechanism.

\subsection{A Tall and Wide View of Missing Data}

Suppose we rearrange the  data  such that the observed ones are ordered first.  Rearrangement is not necessary in practice but it makes the  idea easier to grasp.   Consider the following example:

{\LARGE\footnotesize
\begin{eqnarray*}
Z_0=
\begin{pmatrix}
z_{11} & z_{12} & z_{13} & z_{14}& z_{15} \\
z_{21} & *     & z_{23} & z_{24}& z_{25} \\
z_{31} & z_{32} & z_{33} & z_{34}& z_{35} \\
z_{41} & z_{42} & z_{43} & z_{44}& z_{45} \\
z_{51} & z_{52} & z_{53} & z_{54}& z_{55} \\
z_{61} & z_{62} & z_{63} & *    & z_{65} \\
*     & z_{72} & z_{73} & z_{74}& z_{75} \\
z_{81} & z_{82} & z_{83} & z_{84}& z_{85} \\
\end{pmatrix} \rightarrow Z_1=
\begin{pmatrix}
z_{13} & z_{15} & z_{11} & z_{12} &  z_{14} \\
z_{23} & z_{25} & z_{21} & *     &  z_{24} \\
z_{33} & z_{35} & z_{31} & z_{32} &  z_{34} \\
z_{43} & z_{45} & z_{41} & z_{42} &  z_{44} \\
z_{53} & z_{55} & z_{51} & z_{52} &  z_{54} \\
z_{63} & z_{65} & z_{61} & z_{62} &  *     \\
z_{73} & z_{75} & *     & z_{72} &  z_{74} \\
z_{83} & z_{85} & z_{81} & z_{82} &  z_{84} \\
\end{pmatrix} \rightarrow Z_2=
\left(
\begin{array}{ll|llll}
z_{13} & z_{15} & z_{11} & z_{12} &  z_{14} \\
z_{33} & z_{35} & z_{31} & z_{32} &  z_{34} \\
z_{43} & z_{45} & z_{41} & z_{42} &  z_{44} \\
z_{53} & z_{55} & z_{51} & z_{52} &  z_{54} \\
z_{83} & z_{85} & z_{81} & z_{82} &  z_{84} \\ \hline
z_{23} & z_{25} & z_{21} & *     &  z_{24} \\
z_{63} & z_{65} & z_{61} & z_{62} &  *     \\
z_{73} & z_{75} & *     & z_{72} &  z_{74} \\
\end{array}
\right)
\end{eqnarray*}
}

The transformation from $Z_0$ to $Z_1$ shuffles the columns so that those  that are observed at all times are ordered first. The transformation from $Z_1$ to $Z_2$ shuffles the rows so that time periods with complete data for all units are ordered first.

\begin{figure}[ht]
\caption{Reorganized Data}
\label{fig:fig1}

\vspace*{.25in}
{\small
\begin{center}
\begin{tikzpicture}
  \matrix[
  matrix of math nodes,
  row sep=.5ex,
  column sep=4ex,
  left delimiter=(,right delimiter=),
  nodes={text width=.75em, text height=1.75ex, text depth=.5ex, align=center}
  ] (m)
  {
    \x  & \x  & \x & \x  & \x & \x &  \x & \x\\
    \x   & \text{bal}  & \x & \x & \text{wide} &\x & \x & \x\\
    \x   & \x   & \x & \x & \x & \x &   \x & \x\\
    \x   & \x   & \x  & \x & \y & \x  & \y & \y\\
    \x   &  \x  & \x  & \y & \ldots & \ldots &\y & \x\\
    \x   &  \text{tall}  & \x  & \y &  \text{miss} &  & \y & \y\\
    \x   &  \x  & \x  & \x & \y & \x &\y & \y \\
    \x   &  \x  & \x  & \x & \y & \y &\x & \y \\
    \x   &  \x  & \x  & \x & \y & \y &\x & \y \\
    \x   &  \x  & \x  & \y & \y & \y &\x & \x \\
    };
    \begin{scope}[on background layer]
           \node[fit=(m-1-1)(m-3-8),
            fill=blue!30, rounded corners] {};
            \node[fit=(m-4-1)(m-10-3), fill=blue!15,rounded corners] {};
            \node[fit=(m-1-1)(m-3-3), fill=blue!50,rounded corners] {};
    \end{scope}
  \end{tikzpicture}

\end{center}
}

\end{figure}
 
 We will use `o' to denote the size of the observed and `m' for the size of the  missing samples, respectively. The northwest block, labeled \textsc{bal}, is  a subpanel  of complete data  of dimension $T_{o}\times N_{o}$. The \wide\; block  extends the \textsc{bal} block in the cross-section dimension to include data of all $N$ units with data for $T_o$ periods. The \tall\;  block extends the \textsc{bal} block in the time dimension to include all $N_{o}$ units with complete time series observations. The southeast block  collects the missing data into a $T_m\times N_m$ matrix where $T_m=T-T_{o}$ and $N_m=N-N_{o}$.    This block, labeled \textsc{miss}, is the sub-block bordered by the rows $T_o+1:T$ and columns $N_o+1:N$.  As drawn, \textsc{miss} is a ``largest possible'' block of missing data since some points in it  are actually observed.

To give some economic content, we can think of the  \tall\; block  as  data for developed countries, the \wide\; block  for  newly developed countries which have complete data over a shorter span, while the \textsc{miss} block consists of data for the less developed countries for which missing data are more prevalent. For financial data, acquisitions and mergers can yield a block structure.  In macroeconomic settings, it is not uncommon for statistical agencies  to stagger the release  of data groups, but bunch the release of series within a group. In other cases,  data may not collected in early years and terminated in later years due to attrition. As will be seen below, missing data also play a role in estimation of treatment effects, and  as discussed in  \citet{abdik},  treatment may be given at the end of the sample, or it  may  be staggered or bunched by design of the experiment. In some of these cases, the assumption of missing at random may be inappropriate.


The estimation issue is that principal components cannot be directly applied when there are missing values.
 If we initialize using estimates from the balanced block, the factor estimate will always be  spanned by factors that are originally in the balanced block. If this block is small in size,  the information loss can be significant. Reorganizing data into four blocks reveals that  we can  make better use of the data to estimate the factors.


Our estimator  is based on the idea that   $(\tilde F_{\text{tall}},\tilde \Lambda_{\text{tall}})$ can be obtained from the \tall\; block,   while
$(\tilde F_{\text{wide}},\tilde \Lambda_{\text{wide}})$ can be obtained from the \wide\; block by APC, hence the acronym \textsc{tall-wide}, or \textsc{tw} for short. Results from our previous work can be used to show that
\begin{align*}
\tilde F_{\text{tall},t} &=  H_{\text{tall}}^\prime F_t^0+o_p(1), \quad \tilde F_{\text{wide},t} =H_{\text{wide}}^\prime F_t^0+o_p(1)\\
\tilde \Lambda_{\text{tall},i}&=  H_{\text{tall}}^{-1} \Lambda_i^0+o_p(1), \quad \tilde \Lambda_{\text{wide},i}=  H_{\text{wide}}^{-1} \Lambda_i^0+o_p(1)
\end{align*}
where  $H_{\text{tall}}$ and
$H_{\text{wide}}$ are unknown rotation matrices. The $o_p(1)$ terms are uniform in $i$ and $t$. To make further progress, some assumptions are needed.  Our working assumptions  are  firstly,  that there is an identifiable rank $r$ component in $X$ and in each of its four  blocks so that there is commonality between blocks, and secondly that the observed blocks  are sufficiently large so that consistent estimation by APC is possible.

\paragraph{Assumption B:}
(a) The conditions in Assumption A hold for the full  $T\times N$ matrix $X$ if it were observed, as well as the four sub-blocks:  \textsc{bal},  \textsc{tall},  \textsc{wide}, and  \textsc{miss}.
(b)  The order conditions  $TN_o> r(T+N_o)$ and $T_oN> r(T_o+N)$  are satisfied for any $N,T,N_o,T_o$. Furthermore,
 $\frac{\sqrt{N}}{\min\{N_o,T_o\}}\rightarrow 0$ and $  \frac{\sqrt{T}}{\min\{N_o,T_o\}}\rightarrow 0$ as $N_o,N\rightarrow\infty$ and $T_o,T\rightarrow\infty$.


\paragraph{Assumption C:} (strong sub-block factor and factor loadings)
\begin{itemize}
\item[a.]
 $ \frac{\Lambda_o^{0\prime }\Lambda_o^0}{N_o}\pconv  \Sigma_{\Lambda,o}>0,
  \frac{\Lambda_m^{0\prime}\Lambda_m^0}{N_m}\pconv  \Sigma_{\Lambda,m}>0,
\frac 1 {\sqrt{\No } } \sum_{i=1}^{\No } \Lambda_i^0 e_{it} \dconv N(0,\Gamma_{ot}),$
\item[b.] $\frac{F_o^{0\prime} F_o^0}{T_o}\pconv \Sigma_{F,o}>0,
\frac{F_m^{0\prime} F_m^0}{T_m}\pconv \Sigma_{F,m}>0,
\frac 1 {\sqrt{T_o} } \sum_{s=1}^{T_o} F_s^0 e_{is}\dconv N(0,\Phi_{oi}).$
\end{itemize}


\paragraph{Assumption D:} (block stationarity) Let $\Sigma_\Lambda$, $\Sigma_F$, $\Gamma_t$ and $\Phi_i$ be defined as  in Assumption A.
\begin{itemize}
\item[a.] $\Sigma_{\Lambda,o}=\Sigma_{\Lambda,m}=\Sigma_\Lambda$,  and $\Gamma_{ot}=\Gamma_t$,
\item[b.]
 $\Sigma_{F,o}= \Sigma_{F,m}=\Sigma_F$, and $\Phi_{oi}=\Phi_i$.
\end{itemize}

Assumption B allows $p_N=N_o/N\rightarrow 0$ and $p_T=T_o/T\rightarrow 0$ as $N_o, T_o\rightarrow \infty$ but requires  an order condition to hold for the \textsc{tall} block, and one for the \textsc{wide} block.  Assumption C requires  the subsample moment matrices for the factor and factor loadings to be positive definite so that the factors from the subsamples are identifiable.
Assumption D  restricts these subsample limits to be the same as the limits for the whole sample. Results without Assumption D will also be stated.

 Though there is no  explicit restriction on the missing pattern, it is implicitly imposed  through the factor model. If the \textsc{miss} block has nothing in common with the \textsc{tall} block, the missing values cannot be predicted.
 For example, if   $N_o$ units are driven by a factor $F_1$ and the remaining $N-N_o$ units are driven by a different factor $F_2$, our method will not help with the imputation. Our method is also not suited for situations when  $N_o$ or $T_o$ is small. We certainly cannot handle the case of $T_o=0$ or $N_o=0$. For example,  if  each $(i,t)$ entry is randomly observed with  probability $p\le \frac{1}{2}$ so that when $N\rightarrow\infty$,  $(1-p^N)^N\rightarrow 1$,  $N_o$ will tend to zero and \textsc{tall} block will not be available for estimation.

\begin{lemma}
  \label{lem:lem1-missing}
  Suppose that  there are $N_{o}$ units with complete data in all $T$ rows, and   $T_{o}$ periods when data are available for all $N$ units. Let $(\tilde F_{\text{tall}},\tilde \Lambda_{\text{tall}})$ and $( \tilde F_{\text{wide}},\tilde\Lambda_{\text{wide}})$ be obtained by applying the APC estimator to the \tall\ and \wide\ blocks respectively.  Let $G_{c}=H_{c}^{-1}$ for $c=\text{tall},\text{wide}$. Suppose that  $N_{o}\rightarrow \infty$ as $N\rightarrow\infty$ and $T_{o}\rightarrow\infty$ as $T\rightarrow\infty$.
 Under Assumptions A-D,
\begin{itemize}
\item[a.]  $\sqrt{ N_o}(\tilde F_{\text{tall},t}- H_{\text{tall} }^\prime F^0_{ t})\dconv  N\Big(0,\mathbb D_r^{-2} \mathbb Q_r \Gamma_t \mathbb Q_r^\prime \mathbb D_r^{-2} \Big)$;\newline
$\sqrt{ T} (\tilde\Lambda_{\text{tall}, i}- G_{\text{tall}}\Lambda_{i}^0)\dconv N \Big(0,(\mathbb Q_r^\prime)^{-1}\Phi_i \mathbb Q_r^{-1} \Big)$;
\item[b.]  $\sqrt{N}(\tilde F_{\text{wide},t}- H_{\text{wide} }^\prime F^0_{ t})\dconv  N\Big(0,\mathbb D_r^{-2} \mathbb Q_r \Gamma_t \mathbb Q_r^\prime \mathbb D_r^{-2} \Big)$;
\newline
$\sqrt{ T_o} (\tilde\Lambda_{\text{wide}, i}- G_{\text{wide}}\Lambda_{i}^0)\dconv N\Big(0,(\mathbb Q_r^\prime)^{-1}\Phi_i \mathbb Q_r^{-1} \Big).$

\end{itemize}
\end{lemma}
Lemma \ref{lem:lem1-missing} is a direct implication of Lemma \ref{lem:bai-03}. The asymptotic variances are the same as defined in Lemma \ref{lem:bai-03};  the convergence rates differ because estimation is no longer based on the full sample.

 Under our maintained assumptions, it immediately follows from Lemma \ref{lem:lem1-missing}  that  \begin{eqnarray*}
   \tilde C_{\text{tall},it}&=& \tilde F_{\text{tall},t}^\prime\tilde\Lambda_{\text{tall},i}=C_{it}+o_p(1)\\
\tilde C_{\text{wide},it}&=& \tilde F_{\text{wide},t}^\prime\tilde\Lambda_{\text{wide},i}=C_{it}+o_p(1).
  \end{eqnarray*}  Obtaining an estimate of  $ C_{it}$ for $(i,t)$ in  \textsc{miss}  requires  a bit more work because  $\tilde F_{\text{tall}}$ and $\tilde \Lambda_{\text{wide}}$ are estimated from different blocks of data.
However, if Assumptions A-D hold,  the data in the block \textsc{bal} will contain information shared by both the \textsc{tall} and \textsc{wide}  and can be used to re-rotate the data. Specifically,  for any $i\in\textsc {bal}$, it holds that $\Lambda_i^0= H_{\text{wide}} \tilde\Lambda_{\text{wide},i}+o_p(1) =  H_{\text{tall}}\tilde\Lambda_{\text{tall},i}+o_p(1)$, and
\[\tilde\Lambda_{\text{tall},i}= ( H_{\text{tall}}^{-1} H_{\text{wide}}) \, \tilde\Lambda_{\text{wide},i} +o_p(1), \quad \quad i=1,2,...N_o.\]
 Define a new  $r\times r$ non-singular rotation matrix
\[  H_{\text{miss}} =  H_{\text{tall}}^{ -1} H_{\text{wide}}.\]
This matrix  can be estimated by regressing the $N_{o}\times r $ matrix $\tilde\Lambda_{\text{tall}}$ on the $ N_{o}\times r$ matrix  $\tilde\Lambda_{\text{wide}}$, which is the  $N_o\times r$ {\em sub-matrix} of $\tilde \Lambda_{\text{wide}}$ associated with the balanced block. As shown in the appendix,
 $  H_{\text{tall}}\tilde H_{\text{miss}} H_{\text{wide}}^{-1}   = I_r +O_p(1/N_o+1/T_o)$.  This suggests the following;
\paragraph{Algorithm TW}

  Let  $\Omega$ be the $T\times N$ matrix that is one in positions when the data are observed, i.e. $\Omega_{it}=1$ if $X_{it}$  is observed and zero otherwise.


\begin{itemize}[noitemsep]

\item[1.]  From  the \textsc{tall} block of $ X$, obtain  $(\tilde F_{\text{tall}},\tilde \Lambda_{\text{tall}})$ by APC where $\tilde F_{\text{tall}}$ is $T\times r$.

\item[2.] From  the \wide\; block of $ X$, obtain  $(\tilde F_{\text{wide}},\tilde \Lambda_{\text{wide}})$ by APC where $\tilde\Lambda_{\text{wide}}$ is $N\times r$.
\item[3.] Let $\tilde C_{\text{miss}}=\tilde F_{\text{tall}} \tilde H_{\text{miss}} \tilde\Lambda_{\text{wide}}^\prime$ where  $\tilde H_{\text{miss}}$ is  obtained by regressing $\tilde\Lambda_{\text{tall}}$ on a submatrix of $\tilde\Lambda_{\text{wide}}$.
\item[4.] Output $
   \tilde X_{it}=X_{it}$ if $\Omega_{it}=1$ and
$X_{it}=  \tilde C_{it}$ if $\Omega_{it}=0.$

\end{itemize}

Steps (1) and (2) imply that a  complete set of estimates of the low rank component can be obtained from   $TN_{o}+T_{o}N> T_{o}\times N_{o}$ data points. When  the number of factors in \textsc{tall} and \textsc{wide} do not coincide,  we  let  $r=\max(r_{\text{tall}},r_{\text{wide}})$ in Step (3).

\section{Properties of  $\tilde C$ and Re-estimation}

This section studies  the properties of $\tilde C_{it}$ which will be used to replace the missing $X_{it}$.   The estimation error can be decomposed into four terms:

{\small
\begin{eqnarray*}
 \tilde C_{\text{miss},it} -C_{\text{miss},it}^0 &=&   \tilde F^\pr_{\text{tall},t}\tilde H_{\text{miss}}\tilde \Lambda_{\text{wide},i}  -F_t^{0\pr} \Lambda_i^{0} \\
 &=& \bigg(\tilde F_{\text{tall},t}- H_{\text{tall}}^\prime F_t^0\bigg)^\prime  \tilde H_{\text{miss}}\bigg(\tilde \Lambda_{\text{wide},i}- H_{\text{wide}}^{-1} \Lambda_i^0\bigg)
 +F_t^{0\prime} H_{\text{tall}} \tilde H_{\text{miss}}
    \bigg(\tilde \Lambda_{\text{wide},i}- H_{\text{wide}}^{-1}\Lambda_i^0\bigg)\\
 &&+ \bigg(\tilde F_{\text{tall},t}- H_{\text{tall}}^\prime F_t^0\bigg)^\prime \tilde H_{\text{miss}} H_{\text{wide}}^{-1}\Lambda_i^0   + F_t^{0\prime}\bigg(I_r+ H_{\text{tall}} \tilde H_{\text{miss}}  H^{-1}_{\text{wide}}-I_r\bigg) \Lambda_i^0- F_t^{0\prime}\Lambda^0_i\\
 &=&O_p\bigg(\frac{1}{\sqrt{N_{o}T_{o}}}\bigg) + O_p\bigg(\frac{1}{\sqrt{T_{o}}}\bigg)+O_p\bigg(\frac{1}{\sqrt{N_{o}}}\bigg)+O_p(\frac 1 {N_o}+\frac 1 {T_o}).
 \end{eqnarray*}
}
\begin{proposition}
\label{prop:prop1-missing}
Under the assumptions of A-D,
\[  \delta_{\mathbb N,\mathbb T}\bigg(\frac{\tilde C_{it}-C^0_{it}}{\sqrt{\tilde{\mathbb V}_{it}(\mathbb N,\mathbb T)}}\bigg)\dconv N(0,1))\]
where   $\tilde{\mathbb V}_{it}(\mathbb N,\mathbb T)$ is an estimate of $\mathbb V_{it}(\mathbb N,\mathbb T)=\frac{\delta^2_{\mathbb N\mathbb T}}{\mathbb N} V_{it}+\frac{\delta_{\mathbb N\mathbb T}^2}{\mathbb T} W_{it}$,
$V_{it}=\Lambda_i^{0\prime} \Sigma_{  \Lambda}^{-1}  \Gamma_t \Sigma_{\Lambda}^{-1}\Lambda^0_i$ and $W_{it}=+
F_t^{0\prime} \Sigma_{F}^{-1}  \Phi_i \Sigma_{F}^{-1}F^0_t$ as  defined in (\ref{eq:eqVW}).

\begin{itemize}
\item[i.] for $(i,t)\in $\tall,  $\delta_{\mathbb N,\mathbb T}=\min(\sqrt{N_o},\sqrt{T})$;
\item[ii.] for $(i,t)\in$ \wide,   $\delta_{\mathbb N,\mathbb T}=\min(\sqrt{N},\sqrt{T_o})$;

\item[iii.] for $(i,t)\in$\textsc{bal},
\begin{itemize}
\item[a.] $\tilde C_{it}=\tilde C_{\tall,it}$ and  $\delta_{\mathbb N,\mathbb T} =\min(\sqrt{N_o},\sqrt{T}),$
if $\min(N_{o},T)>\min(N,T_{o})$;
\item[b.] $\tilde C_{it}=\tilde C_{\wide,it}$ and
$\delta_{\mathbb N,\mathbb T}=\min(\sqrt{N},\sqrt{T_o}),$ if $\min(N_{o},T)\le \min(N,T_{o})$;
\end{itemize}
\item[iv.] for $(i,t)\in$ \textsc{miss}, $\delta_{\mathbb N,\mathbb T}=\min(\sqrt{N_o},\sqrt{T_o})$.

\end{itemize}
\end{proposition}

Proposition \ref{prop:prop1-missing} shows that the estimates of the entire $C_{it}$ matrix are consistent and asymptotically normal without explicit restrictions on the nature of  missingness. However, the   convergence rate of $\tilde C_{it}$  depends on whether  $X_{it}$ is observed. Note that
 the   \textsc{bal} block can use  estimates from \tall \; or from  \wide  \; block, so  convergence rate for  this block is the faster of the two rates, ie.
\[\max\bigg(\min(\sqrt{N_o},\sqrt{T}),\min(\sqrt{N},\sqrt{T_o})\bigg).\]
In contrast, the convergence rate of $C_{it}$ in the \textsc{miss} block is always the slowest possible. These convergence rates and asymptotic distributions are obtained without  iteration.   It is  possible for other estimators to achieve a better convergence rate. But to our knowledge, few (if any) consistent estimator exists that does not require iteration. Algorithm \textsc{tw} is best suited for cases when a large \textsc{tall} blow is availble, and the number of missing values is similar across units. In the event that a few units have many more missing values, $T_o$ will be the smallest sample size possible and could result in information loss.  In \citet{bcn:21}, we propose an alternative algorithm  that estimates the loadings and the rotation matrix jointly by  a series of projections.

\subsection{The Imputed Values}

 While  Algorithm \textsc{tw} produces  factor estimates that are mutually orthogonal within the four blocks of data, they are not mutually orthogonal over the entire data matrix.   Furthermore, the estimates in \textsc{tall} do not use all information available, and similarly for the estimates in \textsc{wide}. Re-estimation  using $\tilde X$  provides an opportunity to use the imputed entries not previously available. However,  embedded in   $\tilde X$  are imputation errors which  will propagate to other blocks in  re-estimation   because the APC is a weighted average  of  $\tilde X$. A formal analysis is needed to determine whether  re-estimation  using $\tilde X$ can be justified.

Since $\tilde X$ was constructed using estimates of  $F$ constructed from the \textsc{tall} block, and of $\Lambda$ constructed from the \textsc{\wide} block,  we partition the matrices as follows:
\[ F^0=\begin{bmatrix} \underbrace{F_o^0}_{T_o\times r} \\ \underbrace{F_m^0}_{T_m\times r} \end{bmatrix}, \quad
 \Lambda^0=\begin{bmatrix} \underbrace{ \Lambda_o^0}_{N_o\times r} \\ \underbrace{\Lambda_m^0}_{N_m\times r} \end{bmatrix}, \quad
 \tF_{\text{tall}}=\begin{bmatrix} \underbrace{\tF_o}_{T_o\times r} \\ \underbrace{\tF_m}_{T_m\times r} \end{bmatrix}, \quad
 \tilde \Lambda_{\text{wide}}=\begin{bmatrix} \underbrace{\tilde \Lambda_o}_{N_o\times r} \\ \underbrace{\tilde \Lambda_m}_{N_m\times r}\end{bmatrix} \]
where  $T=T_o+ T_m$ and  $N=N_o+N_m$. We
show in the Appendix that if $X_{it}$ is missing,
\begin{eqnarray}
  \label{main-cit-cit} \tilde C_{it}-C_{it}^0&=& u_{it}+v_{it}+r_{it}
\end{eqnarray}
where $r_{it}=O_p(\delta^{-2}_{T_o,N_o})$ uniformly in $(i,t)$ and
\begin{eqnarray*}
u_{it}&=& \Lambda_i^{0\prime}\bigg(\frac{\Lambdaop_o\Lambda_o^0}{N_o}\bigg)^{-1}\frac 1 {N_o} \sum_{k=1}^{N_o} \Lambda_k^0 e_{kt}=O_p\bigg(\frac{1}{\sqrt{N_o}}\bigg)\\
v_{it}&=&F_t^{0\prime} \bigg(\frac{F_o^{0\prime} F_o^0}{T_o}\bigg)^{-1}\frac 1 {T_o} \sum_{s=1}^{T_o} F_s^0 e_{is}=O_p\bigg(\frac{1}{\sqrt{T_o}}\bigg).
\end{eqnarray*}
 The dependence of these elements on $(T_o,N_o)$ is suppressed to simplify notation.  Now
\begin{align*}
\tilde   X_{it}&=  \Lambda_i^{0\prime}F_t^0 +e_{it} , \quad  & \text{if} \quad \Omega_{it}=1 \\
 \tilde X_{it}&= \Lambda_i^{0\prime}F_t^0 + u_{it}+v_{it} +r_{it}\quad & \text{if} \quad \Omega_{it}=0.
\end{align*}
 Imputation injects three errors into $\tilde X_{it}$ when $X_{it}$ is not observed:- a quantity $r_{it}$ that is negligible, an error from estimating $F_t$, and one from estimating $\Lambda_i$. As a consequence,  $u_{it}+v_{it}+r_{it}$ will differ from the true error $ e_{it}$.

\begin{lemma} \label{lem:averageRate}
For $\tilde X=\tilde U\tilde D\tilde V\pr$, let  $(\hF,\hL)=(\sqrt{T} \tilde U_r,\sqrt{N} \tilde V_r \tilde D_r)$
be  the APC estimates  based on $\tilde X$  with the normalization $\frac{\tilde F^{+\prime}\hF}{T}=I_r$. Let  $H^+=(\Lambdaop\Lambda^0/N)(\Fop\hF/T)\tilde D_r^{-2}$. Then, under Assumptions A and B,
 \begin{itemize}
\item[i.] $ \frac 1 T \sum_{t=1}^T\|\hF_t-H^{+\prime} F^0_t\|^2 =O_p(\delta^{-2}_{N_o,T_o}). $
\item[ii.] $\tilde D_r^2\pconv \mathbb D^2_r$ and $\frac {\Fop\hF}T\pconv  \mathbb Q_r$, where $\mathbb D_r$ and $\mathbb Q_r$ are defined  in  Lemma \ref{lem:bai-03} for  complete data.
 \end{itemize}
\end{lemma}
  \citet{baing-ecta:02} shows that in the complete data case, $\frac{1}{T}\sum_{t=1}^T \|\tilde F_t-H^\prime F_t^0\|^2=O_p(\delta_{NT}^{-2})$. Lemma \ref{lem:averageRate} says that  when the
 factors are estimated from $\tilde X$,  the  convergence rate  depends on the size of the balanced panel $T_o$ and $N_o$, which is evidently slower than  when all data are observed.

To obtain a distribution theory for the factor estimates,    we also need the representation for $\hF$ and $\hL$. We show in the Appendix that
\begin{align*}
    \hF_t-H^{+\pr} F_t^0 &= \begin{cases}  \tilde D_r^{-2} \bigg(\frac{\hFp F^0}T\bigg) \frac 1 N \sum_{i=1}^N \Lambda^0_i e_{it} +\hat \xi_{NT,t} \quad  & t\le T_o \\
  \tilde D_r^{-2}\bigg(\frac{ \hFp F^0}{T}\bigg) \frac 1 N\bigg( \sum_{i=1}^{N_o} \Lambda^0_i e_{it}+\sum_{i=N_o+1}^N \Lambda^0_i(u_{it}+v_{it})\bigg)
    +\hat \xi_{NT,t} \quad & t> T_o\end{cases}
  \end{align*}
where $\hat \xi_{NT,t}=O_p(\delta^{-2}_{N_o,T_o})$ uniformly in $t$.
  The first representation is for those estimates of $F_t$ when $t\le T_o$. Except for the $\hat \xi_{NT,t}$   term that is asymptotically negligible,  the representation is the same as the case when all  data are observed.  More interesting is the  $t>T_o$ case when  $\tilde X_{it}$ has imputation error.   In appendix,
  we show
\[  \frac 1 N \sum_{i=N_o+1}^N \Lambda^0_i(u_{it}+v_{it})= \Big[\frac {\Nm } N \Big(  \frac{\Lambda_m^{0\prime}\Lambda_m^0 }{\Nm} \Big)  \Big( \frac{\Lambda_o'\Lambda_o}{\No} \Big)^{-1}\Big] \frac 1 {\No } \sum_{k=1}^{\No } \Lambda_k^0 e_{kt} +O_p((N\To)^{-1/2}). \]
This means that for $t>T_o$, 
   \[ \tilde F^+_t-H^{+\prime} F_t^0  ={\tilde D_r}^{-2} (\tilde F^{+\prime} F^0/T) {\mathbf B_\Lambda} \frac 1 {\No } \sum_{i=1}^{\No } \Lambda_i^0 e_{it} + \hat \xi_{NT,t}+ O_p((N\To)^{-1/2}) \]
where 
the $r\times r$ matrix ${\mathbf B_\Lambda}$ is defined as
\[ {\mathbf B_\Lambda}= \frac {\No }  N I_r +\frac {\Nm} N \Big(  \frac{\Lambda_m^{0\prime}\Lambda_m^0 }{\Nm} \Big)  \Big( \frac{\Lambda_o'\Lambda_o}{\No} \Big)^{-1}. \]
Thus the convergence rate for $\tilde F^+_t-H' F_t^0 $ is $\sqrt{N_o}$.
Similar derivations show that
  \begin{align*}
    \hL_i- G^+\Lambda_i^0&=\begin{cases}
       H^+\pr \frac{1}{T}\sum_{t=1}^T F_t^0 e_{it}+\hat \eta_{NT,i}  \quad &i \le N_o\\ \\
     H^+\pr \mathbf B_F \frac{1}{T_o}\sum_{t=1}^{T_o} F_t^0e_{it} + \hat\eta_{NT,i}+O_p( (N_o T)^{-1/2}) \quad & i> N_o.\end{cases}
    \end{align*}
    where $G^+=(H^{+})^{-1}$, $\hat\eta_{NT,i} =O_p(\delta_{N_o,T_o}^{-2})$ uniformly in $i$ and
    \[ \mathbf B_F =\frac{\To} T I_r+ \frac {\Tm} T \Big(\frac{F_m^{0\prime}F_m^0}{T_m} \Big) \Big(\frac{F_o^{0\prime} F_o^0}{\To}\Big)^{-1}. \]
Thus the convergence rate for $\hL_i- G^+\Lambda_i^0$ is $\sqrt{T}$ for $i\le N_o$ and the rate is $\sqrt{T_o}$ for $i>N_o$. Note that under Assumption D,
\begin{equation}\label{BL-BF}
\mathbf B_\Lambda \pconv  I_r, \quad \mathbf B_F \pconv I_r
\end{equation}

Given the asymptotic representations, it is relatively easy to derive the limiting distributions.

\begin{proposition}
\label{prop:hatX-coro1}
  Under Assumptions $A-D$, the following holds as $N\rightarrow\infty$ and $T\rightarrow\infty$, with  $G^+=(H^{+})^{-1}$.
\begin{itemize}
  \item[a.] For $t\le T_o$: $ \sqrt{N}(\hF_t -H^{+\prime} F_t^0) \dconv N(0, \mathbb D_r^{-2} \mathbb Q_r \Gamma_t \mathbb Q_r' \mathbb D_r^{-2})$;
\item[b.] For $t>T_o$:
$ \sqrt{N_o}(\hF_t -H^+\pr F_t^0) \dconv N(0,\mathbb D_r^{-2} \mathbb Q_r \Gamma_{t}   \mathbb Q_r' \mathbb D_r^{-2})$;
\item[c.] For $i\le N_o$: $\sqrt{T}(\hL_i-G^+ \Lambda_i^0)\dconv N(0,
\mathbb Q_r^{\prime -1} \Phi_{i}  \mathbb Q_r^{-1})$;
\item[d.] For $i> N_o$: $\sqrt{T_o}(\hL_i-G^+ \Lambda_i^0)\dconv N(0,
\mathbb Q_r^{\prime -1}\Phi_{i}  \mathbb Q_r^{-1})$.
\end{itemize}
\end{proposition}
The main implication of the proposition is that if  $X_{it}$ is  in the balanced block, then $\hF_t$ is $\sqrt{N}$ consistent  while $\hL_i$ is $\sqrt{T}$ consistent. These are faster rates than those stated in Proposition \ref{prop:prop1-missing} without re-estimation. Note that   there is  only a single rotation matrix for the factor estimates (instead of one for \textsc{tall} and one for \textsc{wide})  which are  mutually orthogonal. This is a consequence of the fact that the factors are now estimated from the entire $\tilde X$ matrix instead of the sub-blocks.

The four convergence rates  in Proposition \ref{prop:hatX-coro1}  are derived under the assumption that no data from the  \textsc{miss}  block are available.  Parts (a) and (c) already have the same rate of convergence as in complete data and are unaffected by this assumption. It can be shown that the convergence rates for parts (b) and (d) will be faster when some observations in the \textsc{miss} block are available.  The assumption that no observations in the \textsc{miss} block is consistent with the usual counterfactual matrix $Y(0)$ in the program evaluation analysis to follow.

\begin{proposition}
  \label{thm:missC}
  Let $\tilde C^+_{it}=\hF_t\pr\hL_i$ be the common component estimated from $\tilde X$ in which missing values of $X$ are replaced by the \textsc{tw} estimates of the common component $C$.
  Under Assumptions $A-D$, it holds that as $N\rightarrow\infty$ and $T\rightarrow\infty$,
  \[ \delta_{\mathbb N, \mathbb T}\bigg( \frac{\tilde C^+_{it}-C^0_{it}}{\sqrt{\tilde{\mathbb V}_{it}(\mathbb N,\mathbb T)}} \bigg)
    \dconv  N(0,1)\]
    where   $\tilde{\mathbb V}_{it}(\mathbb N,\mathbb T)$ consistently estimates   $\mathbb V_{it}(\mathbb N,\mathbb T)$ defined in (\ref{eq:eqVW}), 
  \begin{align*}
     \delta_{\mathbb N,\mathbb T}&=\begin{cases} \min(\sqrt{N},\sqrt{T})  \quad\quad & i\le N_o,t\le T_o \\
      \min(\sqrt{N_o},\sqrt{T})  \quad\quad  &  i\le N_o, t> T_o  \\
      \min(\sqrt{N},\sqrt{T_o})
     \quad\quad\quad & i> N_o, t\le T_o\\
      \min(\sqrt{N_o},\sqrt{T_o}) \quad\quad  & i>N_o,t>T_o,
\end{cases}
  \end{align*}

\end{proposition}

The highlight of  Proposition \ref{thm:missC} is that three of  the convergence
rates  are the same as in Proposition \ref{prop:prop1-missing}, but if $X_{it}$ is in \textsc{bal}, the convergence rate   is now $\min(N,T)$, the same as when $X$ were completely observed.  This improvement  is due to the simple fact that $\tilde C_{it}$ is based on data in the \textsc{tall} and \textsc{wide} blocks only, while $\tilde C^+_{it}$ also also exploits information in the \textsc{miss} block. \citet{baing-joe:19}  considers a robust  principal components (\textsc{rpc}) estimator
 $ (\hat F,\hat\Lambda)=
     ( \tilde  F(D^\gamma_r)^{1/2} ,\tilde \Lambda (D^\gamma_r)^{-1/2})$ where $D_{ii}^\gamma=(D_ii-\gamma)_+$, $\gamma>0$ is a regualrization parameter.
If we  replace the \textsc{apc} part of Algorithm \textsc{tw}  by \textsc{rpc},    $\bar C_{it}$ will also have  the four convergence rates, unaffected by regularization.

 From  the proof of Proposition \ref{thm:missC} in Appendix, the asymptotic representation for $\tilde C_{it}^+-C_{it}^0$
 implies the following error average rate in Frobenius norm (denoted $\|\cdot\|$) for the four  blocks:
\begin{itemize}
\item[1.] For  the block defined by $i\le N_o,T\le T_o$:
$ \frac {\|\tilde C_1^+-C_{1}^0\|} {\sqrt{N_o T_o}} =O_p(\frac 1{\sqrt{N}})+O_p(\frac 1{ \sqrt{T}})+ O_p(\delta_{N_o,T_o}^{-2})$.
\item[2.] For the block defined by $i\le N_o, t> T_o$:
$ \frac {\|\tilde C_2^+-C_2^0\|} {\sqrt{N_o T_m}} =O_p(\frac 1{\sqrt{N_o}})+O_p(\frac 1{ \sqrt{T}})+ O_p(\delta_{N_o,T_o}^{-2})$.
\item[3.] For block defined by $i>N_o, T\le T_o$:
$ \frac {\|\tilde C_3^+-C_3^0\|} {\sqrt{N_m T_o}} =O_p(\frac 1{\sqrt{N}})+O_p(\frac 1{ \sqrt{T_o}})+ O_p(\delta_{N_o,T_o}^{-2})$.
\item[4.] For the block defined by $i>N_o,t>T_o$:
$ \frac {\|\tilde C_4^+-C_4^0\|} {\sqrt{N_m T_m}} =O_p(\frac 1{\sqrt{N_o}})+O_p(\frac 1{ \sqrt{T_o}})+ O_p(\delta_{N_o,T_o}^{-2}) $.
  \end{itemize}

This in turn implies an average squared error for the entire common components matrix
\[ \frac {\|\tilde C^+-C^0\|^2} {N T } =\Big[w_1 O_p(\frac 1 N +\frac 1 T)
+w_2 O_p(\frac 1 {N_o}+\frac 1 T )+w_3 O_p(\frac 1 N+\frac 1 {T_o}) +w_4 O_p(\frac 1 {N_m} +\frac 1 {T_m})  \Big]
 +O_p(\delta_{N_o,T_o}^{-4}) \]
 where the weights are  the proportions of block size:
\[ w_1 =\Big(\frac {N_o T_o}{NT}\Big), \, w_2=\Big(\frac {N_o T_m}{NT}\Big),
w_3 = \Big(\frac {N_m T_o}{NT}\Big), \, w_4=\Big(\frac {N_m T_m}{NT}\Big).  \]
The sum of the first four terms in the average squared error  is  $O_p(\frac{1}{N})+O_p(\frac{1}{T})+(1-p_N)(1-p_T) O_p(\frac{1}{N_o}+\frac{1}{T_o})$
where $p_N= N_o/N$ and $p_T=T_o/T$. We have the following.
\begin{corollary}
  \label{cor:avgC}
  Under Assumptions $A$ and $B$:
 \begin{equation*}
 \frac {\|\tilde C^+-C^0\|} {\sqrt{N T }} =\bigg[O_ p\bigg(\frac 1 {\sqrt{N}}\bigg)+  O_p\bigg(\frac 1 {\sqrt{T}} \bigg)\bigg]+\sqrt{(1-p_T)(1-p_N)} \bigg[O_p\bigg( \frac 1 {\sqrt{N_o}}\bigg)+O_p\bigg(\frac 1 {\sqrt{T_o}}\bigg)\bigg].
 \label{tildeC-C-norm}
 \end{equation*}
\end{corollary}

The  first term in the square bracket is present even in the complete data case. The second term
is due entirely to missing data, and the magnitude
 depends on the fraction of  missing data but does not depend on the missing data mechanism.

\noindent
{\bf Remark 1:}
In the machine learning literature, matrix completion problems are typically solved by nuclear norm regularization, and algorithmic errors bounds are  given for fixed $N$ and $T$. \citet{abdik} finds that for   $\sigma$-sub-Gaussian data,  the worse case  bound for the average  error  $\tilde C$ (analogous to Corollary \ref{cor:avgC}) depends on the regularization parameter and the unspecified distribution that generates $\Omega$.  Our analysis complements   the algorithmic error with a theory for asymptotic inference  and   shows that regularization is not needed for consistent estimation of  $C$.  In fact,  we are able to characterize the sampling error of each $\tilde C_{it}$, not just the average over over $i$ and $t$,  while allowing  $NT >> N_{o} T_{o}$. This is made possible by more fully exploiting the factor structure.

 Compared to \citet{su-missing} and \citet{xiong-pelger:19}, the missing pattern plays a less important  role in our analysis because we look at the problem from the perspective of  re-arranged data.   Our results also differ  in that our first step  estimate  is already consistent and asymptotically normal. Re-estimation in  our setup accelerates the convergence rate of a sub-block rather than restores asymptotic normality. Furthermore,  we do not need to assume that the number of missing data points as a fraction of $T\cdot N$ is  bounded away from zero. If the balanced block is of dimension $T_o\times N_o$,  such an assumption would  have implicitly required that $T_o$ and $T$  are of the comparable order, and likewise for $N_o$ and $N$.

\paragraph{Remark 2:} The above results are obtained assuming  block stationarity.  Relaxing Assumption  D will come  at the cost of notation as we will  need matrices $\mathbb D_c,\mathbb Q_c$ ($c$=\tall,\wide), as well as $\Gamma_{ot}$ and $\Phi_{oi}$ for the sub-samples. The following holds without Assumption D, proved in Propositions A.1  A.2, and Corollaries A.1 and A.2 in the Appendix.

\begin{enumerate}
\item  Lemma \ref{lem:lem1-missing}:  the convergence rates are the same, but
\begin{itemize}
\item[i.] for part a, $(\mathbb D_r, \mathbb Q_r, \Gamma_t$) is replaced by $(\mathbb D_{\tall,r}, \mathbb Q_{\tall, r},\Gamma_{ot}$),
$\Phi_i$ remains the same.

\item[ii.]  for part b,   $(\mathbb D_r, \mathbb Q_r, \Phi_i$) is replaced by $(\mathbb D_{\wide,r}, \mathbb Q_{\wide, r},\Phi_{oi}$),
$\Gamma_t$ remains the same
\end{itemize}
where $\mathbb D_{\tall,r}$ and $\mathbb Q_{\tall,r}$ are similarly defined as in  Lemma \ref{lem:bai-03} from
$\Sigma_{\Lambda,o}^{1/2}\Sigma_F \Sigma_{\Lambda,o}^{1/2}$, and  $\mathbb D_{\wide,r}$ and $\mathbb Q_{\wide,r}$ are defined from
$\Sigma_{\Lambda}^{1/2}\Sigma_{F,o} \Sigma_{\Lambda}^{1/2}$.
\item Proposition \ref{prop:prop1-missing}: the convergence rates are the same, but for part i, replace $(\Sigma_\Lambda, \Gamma_t)$ by
$(\Sigma_{\Lambda,o}, \Gamma_{ot})$ in $V_{it}$; for part ii, replace $(\Sigma_{F},\Phi_i$) by $(\Sigma_{F,o},\Phi_{oi})$ in $W_{it}$;  and for part iv, replace
  $(\Sigma_\Lambda,\Sigma_{F},\Gamma_t,\Phi_i$) by $(\Sigma_{\Lambda,o},\Sigma_{F,o},\Gamma_{ot},\Phi_{oi}$).
  
\item Proposition \ref{prop:hatX-coro1}:
 parts $a$ and $c$  remain the same, the other two parts will read
\begin{itemize}
\item[b.] for
 $t>T_o$, $ \sqrt{N_o}(\hF_t -H^+\pr F_t^0) \dconv N(0,\mathbb D_r^{-2} \mathbb Q_r \mathbb B_\Lambda \Gamma_{ot}  \mathbb B_\Lambda' \mathbb Q_r' \mathbb D_r^{-2})$,
\item[d.]
for $i> N_o$, $\sqrt{T_o}(\hL_i-G^+ \Lambda_i^0)\dconv N(0,
\mathbb Q_r^{\prime -1} \mathbb B_F  \Phi_{oi} \mathbb B_F' \mathbb Q_r^{-1})$.
\end{itemize}
 where $\Gamma_{ot}$ and $\Phi_{oi}$ are given in Assumption C, and
\begin{equation}\label{BLBF-limit}
\mathbb B_\Lambda= p_1 I_r+(1-p_1) \Sigma_{\Lambda,m}\Sigma_{\Lambda,o}^{-1}, \quad
\mathbb B_F=p_2 I_r +(1-p_2)\Sigma_{F,m}\Sigma_{F,o}^{-1},
\end{equation}
  are the limits of $\mathbf B_\Lambda$ and $\mathbf B_F$, respectively,
$p_1$ is  the limit of $N_o/N$ and $p_2$ is the limit of $T_o/T$.

\item Proposition \ref{thm:missC}:  $\mathbb V_{it}({\mathbb N,\mathbb T})$  becomes
\[   \mathbb V_{it}({\mathbb N,\mathbb T})=\begin{cases} \frac {\delta_{  N,  T}^2} {  N} V_{it} +
     \frac {\delta_{  N,  T}^2} {  T} W_{it}   \quad\quad & i\le N_o,t\le T_o\\
    \frac {\delta_{  N_o,  T}^2} {  N_o} V_{it}^o +
     \frac {\delta_{  N_o,  T}^2} {  T} W_{it}
     \quad\quad  &  i\le N_o, t> T_o\\
  \frac {\delta_{  N,  T_o}^2} {  N} V_{it} +
     \frac {\delta_{  N,  T_o}^2} {  T_o} W_{it}^o
     \quad\quad & i> N_o,T\le T_o\\
  \frac {\delta_{  N_o,  T_o}^2} {  N_o} V_{it}^o +
     \frac {\delta_{  N_o,  T_o}^2} {  T_o} W_{it}^o     \quad\quad  & i>N_o,t>T_o
     \end{cases} \]
where $ V_{it}$ and $W_{it}$ are the full sample matrices defined in (\ref{eq:eqVW}), $ V_{it}^o= \Lambda_i^{0\prime} \Sigma_{  \Lambda}^{-1} \mathbb B_\Lambda \Gamma_{ot} \mathbb B_\Lambda' \Sigma_{\Lambda}^{-1}\Lambda^0_i,$ and $ W_{it}^o=F_t^{0\prime} \Sigma_{F}^{-1} \mathbb B_F \Phi_{oi}
\mathbb B_F' \Sigma_{F}^{-1}F^0_t$.
\end{enumerate}

\subsection{Finite Sample Properties}

Simulations are used to compare the performance of \textsc{tw} with and without updating.  For comparison, we also consider an iterative EM algorithm considered in  \citet{stock-watson:handbook-16},  which will be denoted \textsc{em}. Using  estimates from the balanced panel as initial values, the algorithm repeatedly regresses $X$ on $F$ and then $X$ on $\Lambda$ by \textsc{ols} till convergence. Note, however, that the converged factor estimates produced by \textsc{em}  may not be  mutually orthogonal.

Data are generated from $F\sim N(0,D_r)$ and $\Lambda\sim N(0,D_r)$ with $r=2$,  the diagonal entries in $D_r$ are equally spaced between 1 and $1/r$, and  $e_{it}\sim N(0,1)$. We report results for $N=T=200$ only as those for $(N,T)=(200,400)$ and $(N,T)=(400,200)$ are similar. For each replication, $\|\tilde C-C^0\|$ is computed for the four blocks. Also reported are results (labeled \textsc{complete}) for  the infeasible case  when all data are observable.

Our theory is silent about how to compute principal components.   In the case of complete data, a common practice is to first standardize the  data. But with missing data, it is unclear whether this is still desirable.   Hence, we consider three versions of the  estimator: one applied  to the standardized data $X$,  one to the demeanend data,  one to the raw data. These are labeled TW(0,1,2) in the tables reported. The mean and standard deviation used in the centering and normalization are computed using the observations available for each series.

Table \ref{tbl:fro}  compares the mean error over 5000 replications, normalized by the size of the corresponding block.   Not surprisingly, the error in estimating the low rank component is inflated by missing data.    As seen from Table \ref{tbl:fro}, re-estimation always reduces the error, and imputation of the raw data (ie. method (2))  always has smaller errors than re-estimation using standardized data (ie. method (0)).  One possible explanation is that when there are few observations from which to estimate  the sample means and standard deviations,   noise could be injected into the factor estimates.  The results for \textsc{em} also favor imputation of the raw data.

 The Frobenius normed error strongly favors $\tilde C(\tilde X)$, but this is based on averaging the error over all $T\times N$ estimates of $C$.  Table \ref{tbl:mse} reports the  root-mean-square-error for four chosen $(i,t)$ pairs, one in each of the four blocks. Evidently,  the estimation error is  largest if $X_{it}$ is in the \textsc{miss} block and smallest when $X_{it}$ is in the \textsc{bal} block.   As in Table \ref{tbl:fro}, the error is smallest when  the factors are re-estimated using $\tilde X$. Both results are consistent with the theory.

 In results not reported, the  squared correlation between $\tilde C_{i}$ and $C_t$ averaged over $i$ is over 0.93  when all data are observed. For the four data points considered in Table \ref{tbl:fro}, the squared correlations are 0.89, 0.85, 0.85, 0.81 using \textsc{TW}, and 0.92, 0.93, 0.90, and 0.86 upon re-estimation. Regardless of  re-estimation, $\tilde C_{it}$ is well approximated by the normal distribution.

 \section{Factor Based Estimation of the Treatment Effects}

If $X$ is  a panel of data on GDP growth, a macroeconomist may be interested in  a counterfactual prediction for some  $i\in[1,N]$ at some time $t^*>T$. The  theory in \citet{baing-ecta:06} can then be used to construct prediction intervals.   We now  show  that Algorithm \textsc{tw} can also be used to  estimate microeconomic type counterfactuals.

Program evaluation is widely used in economic analysis.   Let $\mathcal T$ denote the treated group
and $\mathcal C$ be the control group that is never exposed to treatment. To conform with the notation in this literature, we now index the treatment group  by 1 and the control group by 0. The group size are  $N_1$ and  $N_0$ respectively  with  $N=N_1+N_0$.
Unit $i$ receives treatment in period $T_{0,i}+1$ and thus $T_{0,i}$ is number of pretreatment periods for unit $i$. In this paper, we assume that  $T_{0i}=T_0$ for all $i$ and thus  $T_1=T-T_0$ is also constant across $i$.

Let $Y_{it}(1)$  be the potential outcome if  individual $i$ receives the treatment, and
$Y_{it}(0)$ be the potential outcome of individual $i$ without treatment in period $t$.
The individual  treatment effect is
\begin{eqnarray*}
   \theta_{it}&=& Y_{it}(1)-Y_{it}(0),\quad  i \in\mathcal T, t> T_0.
\end{eqnarray*}
 The sample average treatment effect on the treated at $t>T_0$ (often referred to as sample $\text{ATT}_t$) is:
\begin{eqnarray*}
\theta_t&=&\frac{1}{N_1}\bigg[\sum_{i\in\mathcal  T}Y_{it}(1)- Y_{it}(0)\bigg]=\frac{1}{N_1}\sum_{i\in\mathcal T}\theta_{it}.
\end{eqnarray*}

A variety of methods have been proposed to estimate these quantities. As discussed in  \citet{abdik},   the unconfoundedness regressions literature tends to use  a  single-treated period to impute  the missing potential outcomes in the last period from the control units with similar lagged outcomes.   The synthetic control literature pioneered in \citet{abadie:03}  uses a weighted average of the control units ( $\sum_{i\in\mathcal C} w_i Y_{it}$) as estimate of the counterfactual outcome of the treated unit in the absence of treatment. The method can be specialized to produce  difference-in-difference estimates under some conditions,   see also \citet{doudchenko-imbens:16}. A  `parallel-trend' condition is needed to ensure that the sample path of the weighted average is parallel to the path of the treated in the absence of treatment.

 \citet{hcw:12} assumes that potential outcome has a factor structure and  considers  estimation when the sample size is too small for estimation of the common factors. Their  two step least squares-based estimator can be understood as replacing the latent factors by the outcome of the control units.  To motivate the use of synthetic control methods in comparative case study research, \citet{adh:10} assumes that the outcome variable is driven by common factors.
  Increasingly,   the synthetic control approach is studied from a factor model perspective.  \citet{gobillon-magnac:16} shows that if the true model is a linear factor model, synthetic controls are equivalent to factor models if the  factor loadings and exogenous covariates for the treated  are in the support of these  variables for the control group.   These conditions  are analogous to requiring that the observed and missing blocks are driven by some  factors in common.

In terms of factor-based estimation of treatment effects, \citet{xu:17} directly estimates  the factors by principal components when $N$ and $T$ are large but the theoretical analysis is incomplete.  \citet{li:18} suggests a procedure to  determine $r$ and provides some asymptotic results for the treatment effect of  a single treated unit in the absence of exogenous covariates. \citet{ass:18} analyses the mean-squared error  of a robust synthetic control procedure.  \citet{xiong-pelger:19} allows the probability of missing data to depend on  observed variables, but requires  the fraction of observed data to bounded away from zero.

We will provide a distribution theory for estimates of the individual and average treatment effect without assuming a  missing data mechanism. Indeed, when potential outcome is assumed to have a factor structure, estimation of  treatment effect is very much related to factor analysis with missing data in which the   outcomes for the treated group had they not been treated are the missing values to be recovered. Let  $x_{it}$ be a $K\times 1$ vector of observed covariates.
 Let $D_{it}$ be the treatment indicator for individual $i$ is treated in period $t$. Then
\begin{eqnarray*}
   Y_{it}&=&D_{it}Y_{it}(1)+(1-D_{it})Y_{it}(0)\\
   &=& \theta_{it} D_{it}+x_{it}^\prime\beta+\Lambda_i^\prime F_t+e_{it}
\end{eqnarray*}
 where $F_t$ is $r\times 1$ vector of latent common factors. This is precisely the interactive fixed effect model developed in  \citet{bai-ecma:09} where   $C_{it}=\Lambda_i^\prime F_t$  is  interactive fixed effect. The covariates $x_{it}$ also helps control for known sources of missingness and allows for more general missing patterns.
We only observe   $Y_{it}(1)$ on the treated and thus need to impute the corresponding counterfactual outcome in the absence of treatment. In the terminology of the previous section,  we now have:
\[ Y(0)= \begin{bmatrix}  Y(0)_{T_0\times N_0}  & Y(0)_{T_0\times N_1} \\
Y(0)_{T_1 \times N_0} &  \textsc{miss}    \end{bmatrix}. \]
\paragraph{Algorithm ATT-TW:}
\begin{itemize}
    \item[1] (IFE): Interactive fixed effect estimation of $\beta$ using observations in the control group. Let $ R$  be $T\times N$ matrix of residuals where $ R_{it}=y_{it}-x_{it}^\prime\hat\beta$
    \item[2] (\textsc{tw}): Estimate $F$  from  $ R_{\text{tall}}$  and  $\Lambda$ from the $ R_{\text{wide}}$; compute
    $\tilde C=\tilde F_{\text{tall}}\tilde H_{miss}\tilde\Lambda_{\text{wide}}\pr$ and   $\tilde C^+=\tilde F^+\tilde\Lambda^+\pr.$
    \item[3] Predict $Y_{\text{miss}}(0)$
    \begin{itemize}
      \item[a.] replace the $(i,t)$th entry in  $Y_{\text{miss}}(0)$ by
    $\hat Y_{it}(0) = x_{it}'\hat \beta + \tilde C_{it}$, OR
     \item[b.] replace the $(i,t)$th entry in  $Y_{\text{miss}}(0)$ by
    $ \hat Y_{it}(0) = x_{it}'\hat \beta + \tilde C_{it}^+.$
    \end{itemize}
    \item[4]  Compute the average treatment effect $\widehat{\theta} _{t}=\frac{1}{N_1}\sum_{i\in \mathcal T}\hat\theta_{it}$.
    \end{itemize}
The  analysis to follow assumes   re-estimation of the factors  and so takes $\hat  C_{it}=\tilde C_{it}^+$ from (3b).
\vspace{-0.05in}
\subsection{The Average Treatment Effect}
 Under the assumed factor structure,  we see that for $(i,t) \in$ \textsc{miss}
\begin{eqnarray*}
Y_{it}(0) - \hat Y_{it}(0)&=&
 x_{it}'(\beta-\hat \beta)+C_{it}-\hat C_{it}+e_{it},
\end{eqnarray*}
There are three errors in the counterfactual $\hat Y_{it}(0)$, one from  estimation of $\beta$,  one from estimation of interactive fixed effects, and  an idiosyncratic noise $e_{it}$.
Since $Y_{it}(1)-\hat Y_{it}(0)=\theta_{it}+x_{it}(\beta-\hat\beta)+C_{it}-\hat C_{it}+e_{it}$, it follows that
\[ \widehat{\theta} _{t} - \theta _{t} =\frac 1 {N_1} \sum_{i\in \mathcal T } x_{it}'(\beta-\hat \beta)+  \frac 1 {N_1} \sum_{i\in \mathcal T }(C_{it}-\hat C_{it})
+\frac 1 {N_1} \sum_{i\in \mathcal T } e_{it}. \]
As
$ \frac 1 {\sqrt{N_1} } \sum_{i\in \mathcal T } e_{it}= O_p(1)$, the convergence rate for $\widehat{\theta} _{t} - \theta _{t}$ is  at most $\sqrt{N_1}$. Now $\beta$ is homogeneous across $i$ and $t$ by assumption, and   from \citet{bai-ecma:09},  $\hat \beta-\beta=O_p(\frac{1}{\sqrt{T_0 N_0}})$. The first term is
thus $O_p(1/\sqrt{T_0 N_0})$ and is dominated.    By
(\ref{Cit+-Cit+}) in Appendix (or equation (\ref{cit-cit}) if Step (3a)  is used)
\[ \frac 1 {N_1} \sum_{i\in \mathcal T }(\hat C_{it}-C_{it})=F_t'\bigg(\frac{F'F}{T}\bigg)^{-1}\mathbf B_F \frac 1 {T_0 N_1} \Big(\sum_{i\in \mathcal T}  \sum_{s=1}^{T_0} F_s e_{is}\Big)
 + \bar \Lambda_{\mathcal T}  \bigg(\frac{\Lambda'\Lambda}N\bigg)^{-1}\mathbf B_\Lambda \frac 1 {N_0 } \sum_{k=1}^{N_0 } \Lambda_k e_{kt} +O_p(\delta_{N_0 ,T_0}^{-2}) \]
 where  $\bar \Lambda_{\mathcal T}=\frac 1 {N_1}\sum_{i\in \mathcal T} \Lambda_i$ is the average of factor loadings in the treatment group. If $N_1$ is large, the first term on the right  is $O_p(1/\sqrt{T_0N_1})$
 which is also dominated. This leads
to the  asymptotic representation for the case when $N_1$ large:
\begin{eqnarray*}
   \widehat{ \theta} _{t} - \theta _{t}& =&- \bar \Lambda_{\mathcal T}'  \bigg(\frac{\Lambda'\Lambda}{N}\bigg)^{-1} \mathbf B_\Lambda \frac 1 {N_0 } \sum_{k=1}^{N_0 } \Lambda_k e_{kt} + \frac 1 {N_1 } \sum_{i\in \mathcal T } e_{it} \\&&
 +O_p\bigg(\frac{1}{\sqrt{N_0 T_0}}\bigg) + O_p\bigg(\frac{1}{\sqrt{T_0 N_1}}\bigg) + O_p(1/\delta_{N_0 ,T_0}^2) .
\end{eqnarray*}
The following shows that $\widehat{\theta}_t$ is asymptotically normal.
\begin{proposition}
  Suppose Assumptions A-D and those in \citet{bai-ecma:09} hold. Then as $N_0,T_0,N_1\rightarrow \infty$,
\label{prop:prop-ATT}
\[   \delta_{N_0,N_1} \bigg( \frac{\widehat{ \theta} _{t} - \theta_{t}}{\sqrt{\mathbb V_{\theta,t}}}\bigg)  \dconv N(0,1)  \]
where $\delta_{N_0,N_1}=\min\{\sqrt{N_0}, \sqrt{N_1}\}$  and
$  \mathbb V_{\theta,t}= \frac {\delta_{N_0,N_1}^2} {N_0}  \,  \bar\Lambda_{\mathcal T}'\Big(\frac{\Lambda'\Lambda}{N}\Big)^{-1}\Gamma_{t} \Big(\frac{\Lambda'\Lambda}{N}\Big)^{-1}\bar \Lambda_{\mathcal T}
+  \frac  {\delta_{N_0,N_1}^2} {N_1} \, \sigma_e^2.$
Replace $\Gamma_{t}$ by $\mathbb B_\Lambda \Gamma_{ot} \mathbb B_\Lambda'$ without Assumption D, where $\mathbb B_\Lambda$ is in (\ref{BLBF-limit}) and $\Gamma_{ot}$ is in Assumption C.
\end{proposition}
The proposition highlights that the convergence rate of
$\widehat{ \theta} _{t} - \theta _{t}$  is  $\min(\sqrt{N_0},\sqrt{N_1})$. Let $\hat\sigma_{\theta,t}=\mathbb V_{\theta,t}/\delta_{N_0,N_1}$.
 When  $\Lambda_i$ and $\sigma^2_e$ are replaced by consistent  estimates, the asymptotic 95\% confidence interval for $\theta_t$ is
$(\widehat{ \theta} _t \pm 1.96 \hat \sigma_{\theta,t})$.
 We can estimate $\sigma^2_e$  from  $ \hat e_{it}=Y_{it}(0)-\hat Y_{it}(0)$ of  the control group. (An alternative is to estimate it from   the control group together with the treatment group prior to the treatment period.)
 Then for $K=\text{dim}(\beta)$,
\[ \hat \sigma_e^2=\frac 1{T N_0-r(T+N_0)+r^2-K} \sum_{i\le N_0}\sum_{t=1}^T \hat e_{it}^2.  \]

The estimation of $\theta_t$  by  \textsc{tw} presented above can be generalized to allow $T_0$ to vary with $i$, as in \citet{xu:17}. This approach was first considered in the unpublished dissertation of \citet{cahan-thesis}, and which we analyze further in \citet{bcn:21}.

\subsection{Treatment Effect on a Single Unit}
Consider now the estimation of  treatment effect on a single unit  $j$ for some $j>N_0$.   Then
\begin{eqnarray*}
   \widehat{ \theta} _{jt} - \theta _{jt} &=&   x_{jt}'(\beta-\hat \beta)
- F_t'\bigg(\frac{F'F}{T}\bigg)^{-1} \mathbf B_F \frac 1 {T_0 } \Big( \sum_{s=1}^{T_0} F_s e_{js}\Big)  \\
&&-  \Lambda_j'  \bigg(\frac{\Lambda'\Lambda}{N}\bigg)^{-1} \mathbf B_\Lambda \frac 1 {N_0 } \sum_{k=1}^{N_0 } \Lambda_k e_{kt}
+  e_{jt}
 +O_p(1/\delta_{N_0 ,T_0}^2) .
\end{eqnarray*}
As before, we can ignore the error $x_{jt}'(\beta-\hat \beta)$.  Imputation error from factor estimation (that is, the second and third terms on the right) are
$O(1/\sqrt{T_0})$ and $O(1/\sqrt{N_0})$, respectively. But unlike $\widehat{\theta}_t$, no  averaging is taken  over $i=N_0+1,\ldots, N$ and, as a consequence,
 $e_{jt}$ now dominates the composite estimation error.  The  distribution of
$\widehat \theta _{jt} - \theta _{jt}$ thus depends on the distribution of $e_{jt}$. If one is willing to assume $e_{jt}$ is identically distributed across $i$ and $t$, its distribution can be estimated using the residuals $\hat e_{it}=Y_{it}(0)-\hat Y_{it}(0).$
 The estimated individual treatment effect has variance
\[ \sigma^2_{\hat\theta_{jt}}=\frac 1 {T_0} F_t'\bigg(\frac{F'F}{T}\bigg)^{-1} \Phi_{j} \bigg(\frac{F'F}{T}\bigg)^{-1} F_t + \frac 1 {N_0} \Lambda_j'\bigg(\frac{\Lambda'\Lambda}{N}\bigg)^{-1} \Gamma_{t} \bigg(\frac{\Lambda'\Lambda}{N}\bigg)^{-1} \Lambda_j
+ \sigma_{et}^2.  \]
Again, $\Phi_j$ can be replaced by $\mathbb B_F \Phi_{oj} \mathbb B_F'$ and  $\Gamma_t$ replaced by $\mathbb B_\Lambda \Gamma_{ot}\mathbb B_\Lambda' $ without block stationarity (Assumption D).
 We estimate $\sigma_{et}^2$ by  $\hat \sigma_{et}^2 =\frac 1 {N-1}\sum_{i, i\ne j }^{N} \hat e_{it}^2$.
If  $e_{jt}$ is assumed  to be normally distributed,  an estimate of its variance suffices for a confidence interval to be constructed as
$ \theta_{jt} \in \Big(\widehat{\theta}_{jt} - 1.96 \hat \sigma_{\hat\theta_{jt}}, \widehat{\theta}_{jt} +  1.96 \hat \sigma_{\hat\theta_{jt}}\Big).$

 It is also of interest to consider the average treatment effect over the treatment period for a single unit, defined as $\theta_j= \frac 1 {T_1}  \sum_{s>T_0} \theta_{js}$. Let $\hat \theta_j=\frac 1 {T_1}  \sum_{s>T_0} \hat \theta_{js}$ and $\bar F=\frac{1}{T_1}\sum_{s>T_0} F_s$. It can be shown that a  result similar to Proposition \ref{prop:prop-ATT} holds:
\[\delta_{T_0, T_1}\Big(\frac{\hat \theta_j-\theta_j}{\sqrt{\mathbb V_{\hat\theta_j}}} \Big)\dconv  N(0,1)\]
  where $\delta_{T_0, T_1}=\min\{\sqrt{T_0},\sqrt{T_1} \}$ and
 $ \mathbb V_{\hat\theta_j}=\frac {\delta_{T_0,T_1}^2} {T_0} \bar F'\Big(\frac{F'F}{T}\Big)^{-1}  \Phi_{j }  \Big(\frac{F'F}{T}\Big)^{-1}\bar F  + \frac  {\delta_{T_0,T_1}^2} {T_1} \, \sigma_{ej}^2.$ And again replace $\Phi_j$ by $\mathbb B_F \Phi_{oj } \mathbb B_F'$ without Assumption D.
 The convergence rate  of $\hat \theta_j-\theta_j$ is $\min\{\sqrt{T_0},\sqrt{T_1}\}$.

Table \ref{tbl:att} reports the bias, root-mean-squared-error, and   the probability that the true treatment effect is within two standard errors of the estimated effect (labeled \textsc{covr}). The first panel evaluates $\hat\theta_{j,t}$  at  $(j,t)=(N_0+1,T_0+5)$, the second panel reports $\widehat{\theta}_t$  at  $t=T_0+5$. The estimation error is larger for $\hat\theta_{j,t}$ than $\hat\theta_{T_0}$ since averaging reduces noise. The error decreases with $N_0$ for given $N_1$, but is quite insensitive to changes in $N_1$ for given $N_0$  because the convergence rate is $\min(N_0,N_1)$.  Except when $N_0$ and $T_0$ are both small, the coverage is quite close to 95\%.

\section{Conclusion}
Missing data is prevalent in empirical work. There is a presumption that iteration is needed to impute missing values, and successful matrix recovery requires solving a regularized problem under  a missing at random assumption. This paper shows that if we are willing to impose a strong factor structure, then the entire low rank component of the data can be consistently estimated by our proposed \textsc{tw} procedure  without iteration or regularization.
The methodology can be used within the potential outcomes framework to estimate the effect of treatment on the treated, and a distribution theory is provided.

\newpage
\begin{table}[ht]
  \caption{Simulations: $\frac{1}{\sqrt{\mathbb N\mathbb T}}\|\hat C-C^0\|$}
   \label{tbl:fro}
\begin{center}

    \begin{tabular}{ll|lll|lll|lll|lll} \hline
      &&(0) & (1) & (2) & (0) & (1) & (2) & (0) & (1) & (2) & (0) & (1) & (2)\\ \hline
case & block
 $(\mathbb N,\mathbb T)$
 & \multicolumn{3}{c|}{COMPLETE}
 & \multicolumn{3}{c|}{TW} &
 \multicolumn{3}{c|}{TW Updated}  &

\multicolumn{3}{c}{EM}  \\ \hline

   full & ( 200,200)  & 0.26 & 0.25 & 0.23 & 0.32 & 0.32 & 0.29 & 0.28 & 0.27 & 0.25 & 0.28 & 0.27 & 0.24 \\
   tall & ( 120,200)  & 0.26 & 0.25 & 0.23 & 0.32 & 0.31 & 0.29 & 0.27 & 0.26 & 0.24 & 0.27 & 0.26 & 0.24 \\
   wide & ( 200,120)  & 0.26 & 0.25 & 0.23 & 0.32 & 0.32 & 0.29 & 0.27 & 0.27 & 0.24 & 0.27 & 0.27 & 0.24 \\
    bal & ( 120,120)  & 0.26 & 0.25 & 0.23 & 0.32 & 0.31 & 0.29 & 0.26 & 0.26 & 0.23 & 0.26 & 0.25 & 0.23 \\
   miss & (  80, 80)  & 0.26 & 0.25 & 0.23 & 0.33 & 0.33 & 0.30 & 0.30 & 0.30 & 0.27 & 0.31 & 0.30 & 0.27 \\
\hline
   full & ( 200,200)  & 0.26 & 0.25 & 0.23 & 0.38 & 0.38 & 0.35 & 0.32 & 0.31 & 0.28 & 0.32 & 0.32 & 0.28 \\
   tall & ( 120,200)  & 0.26 & 0.25 & 0.23 & 0.37 & 0.36 & 0.35 & 0.28 & 0.27 & 0.25 & 0.28 & 0.27 & 0.25 \\
   wide & ( 200, 60)  & 0.26 & 0.25 & 0.23 & 0.38 & 0.37 & 0.34 & 0.31 & 0.30 & 0.27 & 0.30 & 0.30 & 0.26 \\
    bal & ( 120, 60)  & 0.26 & 0.25 & 0.23 & 0.37 & 0.36 & 0.34 & 0.28 & 0.26 & 0.24 & 0.26 & 0.26 & 0.23 \\
   miss & (  80,140)  & 0.26 & 0.25 & 0.23 & 0.41 & 0.40 & 0.35 & 0.38 & 0.37 & 0.33 & 0.38 & 0.38 & 0.33 \\
\hline
   full & ( 200,200)  & 0.26 & 0.25 & 0.23 & 0.40 & 0.39 & 0.36 & 0.32 & 0.31 & 0.28 & 0.32 & 0.31 & 0.28 \\
   tall & (  60,200)  & 0.26 & 0.25 & 0.23 & 0.38 & 0.37 & 0.36 & 0.30 & 0.29 & 0.27 & 0.29 & 0.28 & 0.26 \\
   wide & ( 200,120)  & 0.26 & 0.25 & 0.23 & 0.40 & 0.38 & 0.36 & 0.29 & 0.28 & 0.25 & 0.29 & 0.28 & 0.25 \\
    bal & (  60,120)  & 0.26 & 0.25 & 0.23 & 0.38 & 0.37 & 0.36 & 0.27 & 0.26 & 0.23 & 0.26 & 0.26 & 0.23 \\
   miss & ( 140, 80)  & 0.26 & 0.25 & 0.23 & 0.41 & 0.39 & 0.37 & 0.37 & 0.36 & 0.34 & 0.37 & 0.36 & 0.33 \\
\hline
   full & ( 200,200)  & 0.26 & 0.25 & 0.23 & 0.46 & 0.45 & 0.41 & 0.40 & 0.40 & 0.35 & 0.41 & 0.40 & 0.35 \\
   tall & (  60,200)  & 0.26 & 0.25 & 0.23 & 0.42 & 0.41 & 0.40 & 0.33 & 0.32 & 0.30 & 0.33 & 0.32 & 0.29 \\
   wide & ( 200, 60)  & 0.26 & 0.25 & 0.23 & 0.45 & 0.44 & 0.40 & 0.36 & 0.35 & 0.30 & 0.35 & 0.34 & 0.29 \\
    bal & (  60, 60)  & 0.26 & 0.25 & 0.23 & 0.42 & 0.41 & 0.40 & 0.29 & 0.27 & 0.25 & 0.27 & 0.27 & 0.23 \\
   miss & ( 140,140)  & 0.26 & 0.25 & 0.23 & 0.47 & 0.46 & 0.41 & 0.45 & 0.44 & 0.39 & 0.46 & 0.45 & 0.40 \\

\hline

\end{tabular}
  \end{center}

   \small{Note: (0) denotes standardized, (1) demeaned, (2) raw data.
\\ DGP: The $T\times N$ data matrix $X$ is generated by  $X=F\Lambda^\prime+e$, $F\sim N(0,D_r)$, $\Lambda\sim (0,D_r)$ with $r=2$, $e\sim N(0,2.5)$, and $\diag(D_r)=[1;.5]$. $(\mathbb N,\mathbb T$) is the number of columns and rows in the block. Four configurations of missing data are considered with case 1 having the smallest \textsc{miss} block and case 4 the largest. Reported are the mean over 5000 replications.}
\end{table}

\begin{table}[ht]
  \caption{Root Mean-Squared-Error at four $(i,t)$ pairs}
  \label{tbl:mse}

\begin{center}
\begin{tabular}{ll|lll|lll|lll|lll} \hline
 &     &(0) & (1) & (2) & (0) & (1) & (2) & (0) & (1) & (2) & (0) & (1) & (2)\\ \hline
case & $(i,t)$ in
 & \multicolumn{3}{c|}{COMPLETE}
 & \multicolumn{3}{c|}{TW} &
 \multicolumn{3}{c|}{TW Updated}  &

\multicolumn{3}{c}{EM}  \\ \hline

  1 &   tall & 0.26 & 0.25 & 0.23 & 0.32 & 0.31 & 0.29 & 0.27 & 0.27 & 0.25 & 0.28 & 0.27 & 0.25 \\
  1 &   wide & 0.26 & 0.26 & 0.24 & 0.33 & 0.33 & 0.30 & 0.29 & 0.29 & 0.26 & 0.29 & 0.29 & 0.26 \\
  1 &    bal & 0.26 & 0.25 & 0.22 & 0.31 & 0.31 & 0.29 & 0.26 & 0.25 & 0.22 & 0.26 & 0.25 & 0.22 \\
  1 &   miss & 0.26 & 0.26 & 0.23 & 0.34 & 0.33 & 0.30 & 0.31 & 0.31 & 0.27 & 0.31 & 0.31 & 0.27 \\
\hline
  2 &   tall & 0.26 & 0.25 & 0.23 & 0.38 & 0.37 & 0.35 & 0.28 & 0.28 & 0.25 & 0.29 & 0.28 & 0.25 \\
  2 &   wide & 0.26 & 0.26 & 0.23 & 0.40 & 0.40 & 0.35 & 0.36 & 0.35 & 0.31 & 0.36 & 0.35 & 0.30 \\
  2 &    bal & 0.26 & 0.25 & 0.23 & 0.37 & 0.36 & 0.35 & 0.28 & 0.26 & 0.23 & 0.26 & 0.26 & 0.23 \\
  2 &   miss & 0.26 & 0.25 & 0.23 & 0.40 & 0.40 & 0.35 & 0.38 & 0.37 & 0.32 & 0.38 & 0.38 & 0.33 \\
\hline
  3 &   tall & 0.26 & 0.25 & 0.22 & 0.38 & 0.37 & 0.35 & 0.33 & 0.32 & 0.30 & 0.32 & 0.31 & 0.28 \\
  3 &   wide & 0.26 & 0.25 & 0.23 & 0.41 & 0.40 & 0.37 & 0.30 & 0.29 & 0.26 & 0.30 & 0.29 & 0.26 \\
  3 &    bal & 0.26 & 0.26 & 0.23 & 0.39 & 0.38 & 0.37 & 0.27 & 0.26 & 0.24 & 0.27 & 0.26 & 0.23 \\
  3 &   miss & 0.26 & 0.26 & 0.23 & 0.41 & 0.40 & 0.37 & 0.36 & 0.35 & 0.32 & 0.35 & 0.34 & 0.31 \\
\hline
  4 &   tall & 0.26 & 0.25 & 0.23 & 0.43 & 0.42 & 0.40 & 0.35 & 0.34 & 0.32 & 0.35 & 0.33 & 0.31 \\
  4 &   wide & 0.26 & 0.25 & 0.23 & 0.47 & 0.46 & 0.41 & 0.39 & 0.38 & 0.32 & 0.38 & 0.37 & 0.32 \\
  4 &    bal & 0.26 & 0.26 & 0.23 & 0.42 & 0.42 & 0.40 & 0.29 & 0.28 & 0.25 & 0.27 & 0.27 & 0.24 \\
  4 &   miss & 0.26 & 0.26 & 0.23 & 0.47 & 0.46 & 0.41 & 0.44 & 0.43 & 0.38 & 0.44 & 0.43 & 0.37 \\

\hline
  \end{tabular}
\end{center}
   \small{Note: (0) denotes standardized, (1) demeaned, (2) raw data.
\\ DGP: The $T\times N$ data matrix $X$ is generated by  $X=F\Lambda^\prime+e$, $F\sim N(0,D_r)$, $\Lambda\sim (0,D_r)$ with $r=2$, $e\sim N(0,2.5)$, and $\diag(D_r)=[1;.5]$. $(\mathbb N,\mathbb T$) is the number of columns and rows in the block. Four configurations of missing data are considered with case 1 having the smallest \textsc{miss} block and case 4 the largest. Reported are the root-mean-squared error over 5000 replications.}

\end{table}

\begin{table}[ht]
\caption{Estimated Treatment Effects: $r=2$}
\label{tbl:att}
  \begin{center}

  \begin{tabular}{rrr|rrr|rrr}
    $N_1$ & $N_0$ & $T_0$ & bias & rmse &  covr & bias & rmse & covr  \\   \hline
     &&& \multicolumn{3}{c}{$\widehat\theta_{it}$} &
     \multicolumn{3}{|c}{$ \widehat\theta_t$}
     \\ \hline

   5 &   40 &   15 & 0.052 & 1.117 & 0.967 & 0.006 & 0.504 & 0.931  \\
   5 &   80 &   15 & 0.047 & 1.114 & 0.966 & 0.017 & 0.470 & 0.948  \\
   5 &  120 &   15 &-0.008 & 1.095 & 0.970 &-0.022 & 0.487 & 0.931  \\
   5 &  200 &   15 &-0.026 & 1.133 & 0.961 & 0.014 & 0.496 & 0.932  \\
   5 &   40 &   30 & 0.030 & 1.094 & 0.970 & 0.019 & 0.472 & 0.937  \\
   5 &   80 &   30 & 0.018 & 1.051 & 0.969 &-0.020 & 0.469 & 0.943  \\
   5 &  120 &   30 &-0.010 & 1.041 & 0.983 &-0.005 & 0.480 & 0.940  \\
   5 &  200 &   30 & 0.015 & 1.056 & 0.974 & 0.005 & 0.481 & 0.948  \\
   5 &   40 &   50 &-0.009 & 1.058 & 0.976 &-0.007 & 0.451 & 0.958  \\
   5 &   80 &   50 &-0.017 & 1.031 & 0.978 &-0.024 & 0.463 & 0.951  \\
   5 &  120 &   50 &-0.011 & 1.013 & 0.971 &-0.014 & 0.463 & 0.948  \\
   5 &  200 &   50 &-0.014 & 1.020 & 0.968 &-0.015 & 0.455 & 0.964  \\
   5 &   40 &  100 &-0.003 & 1.023 & 0.980 &-0.012 & 0.471 & 0.949  \\
   5 &   80 &  100 &-0.009 & 0.991 & 0.986 & 0.010 & 0.457 & 0.966  \\
   5 &  120 &  100 & 0.032 & 1.036 & 0.975 &-0.002 & 0.455 & 0.956  \\
   5 &  200 &  100 & 0.060 & 0.993 & 0.984 & 0.021 & 0.452 & 0.953  \\ \hline
  20 &   40 &   15 &-0.018 & 1.106 & 0.973 & 0.012 & 0.298 & 0.883  \\
  20 &   80 &   15 & 0.036 & 1.101 & 0.969 &-0.011 & 0.264 & 0.917  \\
  20 &  120 &   15 &-0.035 & 1.106 & 0.970 &-0.001 & 0.264 & 0.909  \\
  20 &  200 &   15 &-0.004 & 1.082 & 0.964 &-0.003 & 0.255 & 0.918  \\
  20 &   40 &   30 & 0.027 & 1.058 & 0.981 &-0.007 & 0.257 & 0.926  \\
  20 &   80 &   30 &-0.079 & 1.071 & 0.967 &-0.005 & 0.245 & 0.937  \\
  20 &  120 &   30 & 0.002 & 1.051 & 0.971 &-0.009 & 0.242 & 0.943  \\
  20 &  200 &   30 &-0.030 & 1.034 & 0.968 &-0.003 & 0.230 & 0.946  \\
  20 &   40 &   50 & 0.035 & 1.028 & 0.983 & 0.006 & 0.251 & 0.938  \\
  20 &   80 &   50 & 0.008 & 1.038 & 0.983 & 0.011 & 0.243 & 0.939  \\
  20 &  120 &   50 & 0.032 & 1.046 & 0.964 &-0.005 & 0.234 & 0.956  \\
  20 &  200 &   50 &-0.004 & 1.026 & 0.974 &-0.006 & 0.225 & 0.958  \\
  20 &   40 &  100 & 0.000 & 1.036 & 0.991 &-0.011 & 0.237 & 0.949  \\
  20 &   80 &  100 & 0.005 & 1.006 & 0.984 & 0.007 & 0.234 & 0.953  \\
  20 &  120 &  100 & 0.065 & 0.988 & 0.978 &-0.006 & 0.232 & 0.944  \\
  20 &  200 &  100 &-0.026 & 1.013 & 0.982 &-0.003 & 0.219 & 0.964  \\

      \hline
  \end{tabular}
\end{center}
{\small Note: ``bias" is the estimation bias; ``rmse" is the root mean square error; ``covr" is the coverage probability.  $\hat\theta_{i,t}$ is evaluated at $(i,t)=(N_0+1,T_0+5)$, and $\hat\theta_t$ is evaluated at $t=T_0+5$.}
\end{table}

\clearpage
\baselineskip=12.0pt
\bibliography{factor,metrics,macro,forecast,metrics2,bigdata}
\newpage

\newcommand{\lambdaUpperCase}{\Lambda}  
\renewcommand{\tL}{\tilde \lambdaUpperCase}
\newcommand{\tLp}{\tilde \lambdaUpperCase^+}
\renewcommand{\hF}{\tilde F}
\newcommand{\tildeF}{\tilde F}
\newcommand{\tildeLam}{\tilde \Lambda}
\renewcommand{\hL}{\tilde \lambdaUpperCase^+}
\newcommand{\breveL}{\breve \lambdaUpperCase^+}
\newcommand{\imputedCit}{\tilde C^+_{it}}

\renewcommand{\calE}{{\cal E}}

\newcommand{\mathbbD}{{\mathbb D}}
\newcommand{\mathbbQ}{{\mathbb Q}}
\renewcommand{\To}{T_o}  
\renewcommand{\Tm}{T_m}    
\renewcommand{\No}{N_o}
\renewcommand{\Nm}{N_m}

\newcommand{\Lambdao}{\Lambda_o}   
\newcommand{\Lambdam}{\Lambda_m}    
\newcommand{\Fo}{F_o}
\newcommand{\Fm}{F_m}
\newcommand{\hFplusp}{\tilde F^{+\prime}}
\newcommand{\hFplus}{\tilde  F^+}

\def\theequation{A.\arabic{equation}}
\setcounter{equation}{0}
\setcounter{section}{0}
\setcounter{lemma}{0}
\setcounter{proposition}{0}
\setcounter{corollary}{0}
\setcounter{theorem}{0}
\def\thesecton{A.\arabic{section}}
\def\thelemma{A.\arabic{lemma}}
\def\thetheorem{A.\arabic{theorem}}
\def\thecorollary{A.\arabic{corollary}}
\def\theproposition{A.\arabic{proposition}}
\section*{Appendix A}

This appendix provides proofs to the results in the main text along with more general results that are of independent interest.
Throughout, $\|A\|$  denotes the Frobenius norm of matrix $A$.
\bigskip

Recall the notation
$ T=\To+ \Tm ; N=\No +\Nm $
and
\[ F^0=\begin{bmatrix} F_o^0 \\ F_m^0 \end{bmatrix}^{T\times r}, \quad
 \Lambda^0=\begin{bmatrix} \Lambda_o^0 \\ \Lambda_m^0 \end{bmatrix}^{N\times r}, \quad
 \tilde F_{tall}=\begin{bmatrix} \tilde F_o \\ \tilde F_m \end{bmatrix}, \quad
 \tilde \Lambda_{wide}=\begin{bmatrix} \tilde \Lambda_o \\ \tilde \Lambda_m\end{bmatrix} \]
where $F_o^0$ is $T_o\times r$, and $F_m^0$ is $T_m\times r$;  $\Lambda_o^0$ is $N_0\times r$ and $\Lambda_m^0$ is $N_m\times r$.
The partitions of $\tilde F_{tall}$ and $\tilde \Lambda_{wide}$ are the same. We write $(i,t)\in \Omega$ if $\Omega_{it}=1$, and
$(i,t)\in \Omega_\bot$ if $\Omega_{it}=0$.


Similar to the rotation matrix $H$ given in Section \ref{sec:section2}, let
$H_{tall}'= D_{tall}^{-2}(\tilde F_{tall}'F^0/T)(\Lambda_o'\Lambda_o/\No )$, then
 the \textsc{tall} estimator satisfies (see  \citet{bai-ecta-03}, Theorem 1)
\begin{equation}\label{FTallt}  \tilde F_{tall,t}-H_{tall}' F_t^0=D_{tall}^{-2}(\tilde F_{tall}'F^0/T) \frac 1 {\No } \sum_{k=1}^{\No } \lambdaUpperCase_k^0 e_{kt}+\xi_{NT,t}, \quad t=1,2,...,T \end{equation}
where $\xi_{NT,t}=O_p(1/\No +1/T)$ uniformly in $t$. Similarly,  there is a rotation matrix $H_{wide}$ such that for each $i$, the \textsc{wide} estimator satisfies (see  \citet{bai-ecta-03}, Theorem 2)
\begin{equation} \label{LWidei} \tL_{wide,i}-H_{wide}^{-1}\lambdaUpperCase_i^0= H_{wide}'\frac 1 {\To} \sum_{s=1}^{\To} F_s^0 e_{is} +\eta_{NT,i}, \quad i=1,2,...,N \end{equation}
where $\eta_{NT,i}=O_p(1/\To+1/N)$ uniformly in $i$.
Let $H_{miss}=H_{tall}^{-1}H_{wide}$,  and define for $(i,t)\in \Omega_\bot$
\[ \tilde C_{it}=\tilde F_{tall,t}' \tilde  H_{miss} \tilde \lambdaUpperCase_{wide,i} \]
where $\tilde H_{miss}$ is an estimator for $H_{miss}$ obtained by regressing $\tilde \Lambda_{tall,i}$ on
$\tilde \Lambda_{wide,i}$ for $i=1,2,...,N_o$.

\begin{lemma} \label{TW-lemma1} Under the assumptions of Proposition \ref{prop:prop1-missing}:\\
(i) \begin{equation}\label{H3-H3} \tilde H_{miss}= H_{tall}^{-1}H_{wide} +O_p(1/\delta_{\No \To}^2)  \end{equation}
(ii) For $(i,t)\in \Omega_\bot$
\begin{equation}\label{cit-cit} \tilde C_{it}-C_{it}^0= \underbrace{\lambdaUpperCase_i^{0\prime} (\Lambdaop_o\Lambda_o^0/\No )^{-1}\frac 1 {\No } \sum_{k=1}^{\No } \lambdaUpperCase_k^0 e_{kt}}_{u_{it}}
+\underbrace{F_t^{0\prime} (F_o^{0\prime} F_o^0/\To)^{-1}\frac 1 {\To} \sum_{s=1}^{\To} F_s^0 e_{is}}_{v_{it}} +r_{NT,it} \end{equation}
where $r_{NT,it}=O_p(1/\To+1/\No )$ uniformly in $(i,t)\in \Omega_\bot$.
\end{lemma}

\paragraph{Proof of Lemma \ref{TW-lemma1}}
Consider part (i).
Rewrite the representation in (\ref{LWidei}) as
\begin{equation} \label{LWidei5}  H_{wide} \tilde \lambdaUpperCase_{wide,i}=\lambdaUpperCase_i^0 +
(F_o^{0\prime} F_o^0/\To)^{-1}\frac 1 {\To} \sum_{s=1}^{\To} F_s^0 e_{is} +
O_p(1/\delta_{N \To }^2).  \end{equation}
This follows from [see Bai (2003), p 166) and Bai and Ng (2019)],
\[  H_{wide} H_{wide}'= (F_o^{0\prime} F_o^0/\To)^{-1} +O_p(1/\delta_{N \To }^2). \]

Similarly, the Tall estimator has  the asymptotic representation
\[  H_{tall} \tilde \lambdaUpperCase_{tall,i}=\lambdaUpperCase_i^0 +
(F^{0\prime} F^0/T)^{-1}\frac 1 {T} \sum_{s=1}^{T} F_s^0 e_{is} +
O_p(1/\delta_{\No  T }^2).  \]
This implies $\tL_{tall,i}=H_{miss} \tL_{wide,i} + \varphi_{NT,i}, \quad i=1,2...,\No$,
where
\[ \varphi_{NT,i}=H_{tall}^{-1}(F^{0\prime} F^0/T)^{-1}\frac 1 {T} \sum_{s=1}^{T} F_s^0 e_{is}-H_{tall}^{-1} (F_o^{0\prime} F_o^0/\To)^{-1}\frac 1 {\To} \sum_{s=1}^{\To} F_s^0 e_{is} +
O_p(1/\delta_{\No  \To }^2).  \]
Regression gives
\[ \tilde H_{miss}=\Big(\frac 1 {\No } \sum_{i=1}^{\No } \tL_{tall,i}\tL_{wide,i}'\Big)
\Big(\frac 1 {\No } \sum_{i=1}^{\No } \tL_{wide,i}\tL_{wide,i}'\Big)^{-1}. \]
It follows that
\[ \tilde H_{miss}=H_{miss} +\Big(\frac 1 {\No } \sum_{i=1}^{\No } \varphi_{NT,i}\tL_{wide,i}'\Big)
\Big(\frac 1 {\No } \sum_{i=1}^{\No } \tL_{wide,i}\tL_{wide,i}'\Big)^{-1}. \]
Note $\frac 1 {\No } \sum_{i=1}^{\No } \tL_{wide,i}\tL_{wide,i}'$ converges in probability to a positive definite matrix.
Consider the numerator,
\[ \frac 1 {\No } \sum_{i=1}^{\No } \varphi_{NT,i}\tL_{wide,i}'
=O_p(1) \frac 1 {\No T} \sum_{i=1}^{\No } \sum_{s=1}^{T} F_s^0 \tL_{wide,i}' e_{is} +
O_p(1) \frac 1 {\No \To} \sum_{i=1}^{\No } \sum_{s=1}^{\To} F_s^0 \tL_{wide,i}' e_{is} +O_p(1/\delta_{\No  \To }^2). \]
Replace $\tL_{wide,i}'$ by $\lambdaUpperCase_i^{0\prime}H_{wide}'$ (ignore higher orders), we see the two terms on the right hand side
are each $O(1/\sqrt{\No \To})$, thus dominated by $O_p(1/\delta_{\No  \To }^2)$. This proves
(\ref{H3-H3}).



We next proof part (ii).
We can rewrite the representations in (\ref{FTallt}) by
\begin{equation}\label{FTallt2}   H_{tall}^{-1\prime}\tilde F_{tall,t}-F_t^0= (\Lambda_o'\Lambda_o/\No )^{-1}\frac 1 {\No } \sum_{k=1}^{\No } \lambdaUpperCase_k e_{kt}
+ O_p(1/\delta_{\No  T}^2). \end{equation}
This follows by multiplying   (\ref{FTallt}) by $ H_{tall}^{-1\prime}$   and using
\begin{eqnarray*} H_{tall}^{-1\prime}  D_{tall}^{-2}(\tilde F_{tall}'F^0/T)& = & H_{tall}^{-1\prime}  D_{tall}^{-2}(\tilde F_{tall}'F^0/T)(\Lambda_o'\Lambda_o/\No )(\Lambda_o'\Lambda_o/\No )^{-1} \\
& =& H_{tall}^{-1\prime} H_{tall}' (\Lambda_o'\Lambda_o/\No )^{-1}=(\Lambda_o'\Lambda_o/\No )^{-1}, \end{eqnarray*}
where $\tilde F_{tall}=(\tilde F_{tall,1},...,\tilde F_{tall,T})'$. The second equality uses the definition of
$H_{tall}'$.
Rewrite (\ref{LWidei5}) as
\begin{equation}\label{LWidei2} H_{wide} \tL_{wide,i}-\lambdaUpperCase_i^0= (F_o^{0\prime} F_o^0/\To)^{-1}\frac 1 {\To} \sum_{s=1}^{\To} F_s^0 e_{is} +
O_p(1/\delta_{N \To }^2).  \end{equation}

Multiply (\ref{FTallt2}) by $\lambdaUpperCase_i^{0\prime}$ and multiply (\ref{LWidei2}) by $\Fop$, we have
\[ \lambdaUpperCase_i^{0\prime} (  H_{tall}^{-1\prime}\tilde F_{tall,t}-F_t^0)  =v_{it}+ O_p(1/\delta_{\No  T}^2), \]
\[ \Fop_t(H_{wide} \tL_{wide,i}-\lambdaUpperCase_i^0) = u_{it} +  O_p(1/\delta_{N \To }^2),  \]
where $u_{it}$ and $v_{it}$ are defined earlier.
Thus
\begin{eqnarray*} \tilde F_{tall,t}'H_{tall}^{-1}H_{wide}\tL_{wide,i}-\Fop_t\lambdaUpperCase_i^0 & =& (H_{tall}^{-1\prime}\tilde F_{tall,t}-F_t^0+F_t^0)'(H_{wide} \tL_{wide,i}-\lambdaUpperCase_i^0+\lambdaUpperCase_i^0)-C_{it}^0 \\
 &= & (H_{tall}^{-1\prime}\tilde F_{tall,t}-F_t^0)'(H_{wide} \tL_{wide,i}-\lambdaUpperCase_i^0) \\
 && +\Fop_t(H_{wide} \tL_{wide,i}-\lambdaUpperCase_i^0)
   + (H_{tall}^{-1\prime}\tilde F_{tall,t}-F_t^0)'\lambdaUpperCase_i^0 \\
 & = &  O_p(1/\sqrt{\No \To})+ u_{it}+ v_{it}+ O_p(1/\delta_{\No \To}^2)\\
 &  = & u_{it}+v_{it}+O_p(1/\delta_{\No \To}^2). \end{eqnarray*}
Using (\ref{H3-H3}),
we have
\begin{eqnarray*}  \tilde  C_{it} -C_{it}^0 & = & \tilde F_{tall,t}' \tilde H_{miss} \tL_{wide,i} -C_{it}^0 \\
 & = & \tilde F_{tall,t}' (\tilde H_{miss}-H_{tall}^{-1}H_{wide})\tL_{wide,i} +
 \tilde F_{tall,t}'H_{tall}^{-1}H_{wide}\tL_{wide,i}-C_{it}^0 \\
 & = & u_{it}+v_{it}+O_p(1/\delta_{\No \To}^2) \end{eqnarray*}
proving (\ref{cit-cit}).

\paragraph{Proof of Lemma \ref{lem:lem1-missing}.}

Lemma \ref{lem:lem1-missing} is obtained by applying  Lemma \ref{lem:bai-03} to the tall and wide blocks separately. We first prove the case without Assumption D (no block stationarity). As in Lemma \ref{lem:bai-03}, the block size determines the rates of convergence.
For the tall block, the dimension is $T\times N_o$ so the rate for the estimated factor loading is $\sqrt{T}$, and the rate for the estimated
factor is $\sqrt{N_o}$.  The asymptotic variances depend on the  population matrices $(\Sigma_{\Lambda,o},\Sigma_F,\Gamma_{ot},\Phi_i)$. The matrices $\mathbb D_{\tall,r}$ and $\mathbb Q_{\tall,r}$ are related to the eigenvalues and  eigenvectors of
$\Sigma_{\Lambda,o}^{1/2}\Sigma_F \Sigma_{\Lambda,o}^{1/2}$, as described in Lemma \ref{lem:bai-03}.  With these notations, part II(i) follows from Lemma \ref{lem:bai-03}.
Under Assumption D (block stationarity), $(\Sigma_{\Lambda,o},\Sigma_F,\Gamma_{ot},\Phi_i)=(\Sigma_\Lambda,\Sigma_F,\Gamma_t,\Phi_i)$,
 so $\mathbb D_{\tall,r}$ and $\mathbb Q_{\tall,r}$ coincide with
$\mathbb D_r$ and $\mathbb Q_r$, and  part II(i) specializes to Part I(a). The argument for the wide block is the same, where the corresponding
populations matrices are $(\Sigma_{\Lambda},\Sigma_{F,o},\Gamma_{t},\Phi_{oi})$, the matrices $\mathbb D_{\wide,r}$ and $\mathbb Q_{\wide,r}$ are related to the eigenvalues and  eigenvectors of the matrix
$\Sigma_{\Lambda}^{1/2}\Sigma_{F,o} \Sigma_{\Lambda}^{1/2}$.  The rest of the argument is the same.

\paragraph{Proof of Proposition \ref{prop:prop1-missing}.}  This proposition is also a direct consequence of Lemma \ref{lem:bai-03}. Applying Lemma \ref{lem:bai-03} to the tall and wide blocks separately, we obtain the results in (i), (ii), and (iii).
For the missing block, that is, for $(i,t)\in \Omega_\bot$, the limiting distribution for the estimated common component follows from the asymptotic representation in (\ref{cit-cit}). By Assumption C,
\[ \sqrt{N_o} \bigg( \frac{ \lambdaUpperCase_i^{0\prime} (\Lambdaop_o\Lambda_o^0/\No )^{-1}\frac 1 {\No } \sum_{k=1}^{\No } \lambdaUpperCase_k^0 e_{kt} }  {\Lambda_i^{0\prime} \Sigma_{\Lambda,o}^{-1}  \Gamma_{ot} \Sigma_{\Lambda,o}^{-1}\Lambda^0_i} \bigg) \dconv N(0,1), \]
\[ \sqrt{T_o} \bigg( \frac{ F_t^{0\prime} (F_o^{0\prime} F_o^0/\To)^{-1}\frac 1 {\To} \sum_{s=1}^{\To} F_s^0 e_{is} }
{F_t^{0\prime} \Sigma_{F,o}^{-1}  \Phi_{oi} \Sigma_{F,o}^{-1}F^0_t} \bigg) \dconv N(0,1). \]
Thus the convergence rate for $\tilde C_{it}-C_{it}^0$ is $\min(\sqrt{N_o},\sqrt{T_o})$.
The expression for $\mathbb V(\mathbb N,\mathbb T)$ is
\[  \mathbb V_{it}(\mathbb N,\mathbb T) =  \frac{\delta^2_{\mathbb N\mathbb T}}{\mathbb N}\Lambda_i^{0\prime} \Sigma_{\Lambda,o}^{-1}  \Gamma_{ot} \Sigma_{\Lambda,o}^{-1}\Lambda^0_i+
     \frac{\delta^2_{\mathbb N\mathbb T}}{\mathbb T} F_t^{0\prime} \Sigma_{F,o}^{-1}  \Phi_{oi} \Sigma_{F,o}^{-1}F^0_t.\]
With block stationarity (Assumption D), the above becomes $\frac{\delta^2_{\mathbb N\mathbb T}}{\mathbb N} V_{it}+ \frac{\delta^2_{\mathbb N\mathbb T}}{\mathbb T} W_{it}$, where $V_{it}$ and $W_{it}$ are defined in (\ref{eq:eqVW}).


\section*{Analysis based on imputed data matrix}
 Let
$\tilde X_{it}=X_{it}, (i,t)\in \Omega $ and
$ \tilde X_{it} = \tilde C_{it}, (i,t)\in \Omega_\bot $
so  the missing values are replaced by the estimated common components $\tilde C_{it}$. We have
$ X_{it}=  \lambdaUpperCase_i^{0\prime}F_t^0 +e_{it} , \quad (i,t) \in \Omega $
and
$ \tilde X_{it}= \lambdaUpperCase_i^{0\prime}F_t^0 + u_{it}+v_{it} +r_{NT,it}, \quad (i,t)\in \Omega_\bot.$
Consider estimating the factor and factor loadings using the $T\times N$  matrix $\tilde X=(\tilde X_{it})$. Let $\tildeF^+$ be the first $r$ eigenvectors corresponding to the first $r$ largest eigenvalues (arranged in decreasing order) of the matrix $\tilde X\tilde X'/(NT)$ with the normalization $\hFplusp \tildeF^+/T=I_r$, that is,
\[ \frac 1 {NT}\tilde X \tilde X' \tildeF^+ =\tildeF^+ {\tilde D_r}^2 \]
where ${\tilde D_r}^2$ is an $r\times r$ diagonal matrix consisting of the eigenvalues. Let $\tildeLam^+ =\frac 1 T \tilde X'\tildeF^+$. Define the rotation matrix
\[ H^+= (\Lambdaop\Lambda^0/N)(\Fop\tildeF^+/T){\tilde D_r}^{-2}. \]
Introduce
  \[  u=(u_{it}), \quad v=(v_{it}), \quad r_{NT}=(r_{NT,it}) \]
   all $\Tm \times \Nm $ matrices, where
 $u_{it}$, $v_{it}$ and $r_{NT,it}$ are defined earlier.
We further put
\[ F^0=\begin{bmatrix} F_o^0 \\ F_m^0 \end{bmatrix}, \quad
 \Lambda^0=\begin{bmatrix} \Lambda_o^0 \\ \Lambda_m^0 \end{bmatrix}, \quad
 e=\begin{bmatrix} \calE_{11} & \calE_{12}\\ \calE_{21} & \calE_{22} \end{bmatrix},\quad
 e^\dag=\begin{bmatrix} \calE_{11} & \calE_{12}\\ \calE_{21} & u+v \end{bmatrix}, \quad
 R_{NT} =\begin{bmatrix} 0 & 0\\ 0 & r_{NT}  \end{bmatrix}\]
 where $\calE_{jk}$ are sub-blocks of $e$, partitioned conformably, for example,
  \[ \calE_{21}=\begin{bmatrix}
e_{\To+1,1} & e_{\To+1,2} & \cdots & e_{\To+1,\No } \\
e_{\To+2,1} & e_{\To+2,2} & \cdots & e_{\Tm +1,\No } \\
 \vdots & \vdots &   &  \vdots &  \\
e_{T,1} & e_{T,2} & \cdots & e_{T,\No }
\end{bmatrix} ^{\Tm  \times \No } =\begin{bmatrix} (\calE_{21})_{\To+1}' \\ (\calE_{21})_{\To+2}'\\ \vdots \\ (\calE_{21})_{T }'\end{bmatrix} \]
with $(\calE_{21})_{T_o+t}'$ representing the $t$th row of $\calE_{21}$ $(t=1,2,...,\Tm )$ with $T=\To+\Tm$.

For notational simplicity, we  use  $F$ and $F^0$ interchangeably (that is, we
 may suppress the superscript ``0" from $F^0$, $F_o^0$, $F_m^0$ and $F_t^0$).  The same is true for $\Lambda$ and $\Lambda^0$. They represent the true quantities.
With these notations, we have
 \[ X = F \Lambda'+e, \text{ and }
  \tilde X =F \Lambda' +e^\dag + R_{NT}\]
Let $B=(\Lambda_o'\Lambda_o/\No )^{-1}$. We can write the matrix $u$ as
\[ u= \frac 1 {\No } \begin{bmatrix}
(\calE_{21})'_{\To+1}\Lambda_o B \lambdaUpperCase_{\No +1} & (\calE_{21})'_{\To+1} \Lambda_o B \lambdaUpperCase_{\No +2} & \cdots & (\calE_{21})'_{\To+1}\Lambda_o B\lambdaUpperCase_N \\
(\calE_{21})'_{\To+2}\Lambda_o B \lambdaUpperCase_{\No +1} & (\calE_{21})'_{\To+2} \Lambda_o B \lambdaUpperCase_{\No +2} & \cdots & (\calE_{21})'_{\To+2}\Lambda_o B\lambdaUpperCase_N \\
 \vdots & \vdots &   &  \vdots &  \\
(\calE_{21})'_{T}\Lambda_o B \lambdaUpperCase_{\No +1} & (\calE_{21})'_{T} \Lambda_o B \lambdaUpperCase_{\No +2} & \cdots & (\calE_{21})'_{T}\Lambda_o B\lambdaUpperCase_N
\end{bmatrix}
 = \frac 1 {\No } \calE_{21} \Lambda_o B \Lambda_m' \]
Let $A=(F_o^{0\prime}F_o^0/\To)^{-1}$, we can write $v$ as
\[ v=\begin{bmatrix}
F^{0\prime}_{\To+1}  A \frac 1 {\To} \sum_{s=1}^{\To}F_s^0 e_{\No +1,s} & F^{0\prime}_{\To+1}   A \frac 1 {\To} \sum_{s=1}^{\To}F_s^0 e_{\No +2,s} & \cdots & F^{0\prime}_{\To+1}  A\frac 1 {\To} \sum_{s=1}^{\To}F_s^0 e_{N,s} \\
F^{0\prime}_{\To+2}  A \frac 1 {\To} \sum_{s=1}^{\To}F_s^0 e_{\No +1} & F^{0\prime}_{\To+2}   A \frac 1 {\To} \sum_{s=1}^{\To}F_s^0 e_{\No +2,s} & \cdots & F^{0\prime}_{\To+2}  A\frac 1 {\To} \sum_{s=1}^{\To}F_s^0 e_{N,s} \\
 \vdots & \vdots &   &  \vdots &  \\
F^{0\prime}_{T}  A \frac 1 {\To} \sum_{s=1}^{\To}F_s^0 e_{\No +1,s} & F^{0\prime}_{T}   A \frac 1 {\To} \sum_{s=1}^{\To}F_s^0 e_{\No +2,s} & \cdots & F^{0\prime}_{T}  A\frac 1 {\To} \sum_{s=1}^{\To}F_s^0 e_{N,s}
\end{bmatrix} \]
\[ = \frac 1 {\To} F_m^0 A F_o^{0\prime} \calE_{12} \]

From $\frac 1 {NT} \tilde X \tilde X' \tildeF^+ =\tildeF^+ {\tilde D_r}^2$,
we have
\[ \frac 1 {NT} (F\Lambda'+e^\dag+R_{NT})(\Lambda F'+e^{\dag\prime}+R_{NT}) \tildeF^+ =\tildeF^+ {\tilde D_r}^2. \]
The terms involving $R_{NT}$ are dominated. We focus on the remaining terms.
Expanding the preceding equation, ignoring the terms involving $R_{NT}$, we obtain
\begin{equation}\label{identity1} F(\Lambda'\Lambda/N)(F'\tildeF^+/T)+F\Lambda'e^{\dag\prime} \tildeF^+/(NT) +e^\dag \Lambda F'\tildeF^+/(NT)+ e^\dag e^{\dag\prime}\tildeF^+/(NT) =\tildeF^+ {\tilde D_r}^2,\end{equation}
\[ \tildeF^+ -F H^+=\Big(F\Lambda'e^{\dag\prime} \tildeF^+/(NT) +e^\dag \Lambda F'\tildeF^+/(NT)+ e^\dag e^{\dag\prime}\tildeF^+/(NT) \Big) {\tilde D_r}^{-2}. \]

\noindent
{\bf Proof of Lemma \ref{lem:averageRate}(i).} We first collect some basic results. Notice
\[ \frac 1 T \| \frac 1 N e\Lambda\| ^2 =
\frac 1 T \sum_{t=1}^T \|\frac 1 N e_t'\Lambda\|^2=\frac 1 T \sum_{t=1}^T \|\frac 1 N \sum_{i=1}^N \lambdaUpperCase_i e_{it}\|^2 =O_p(1/N), \]
where $e_t'$ is the $t$th row of matrix $e$  (or $e'=(e_1,e_2,...,e_T)$.)
Similarly,
\begin{equation} \label{1stBlock} \frac 1 {\To} \|\frac 1 {\No } \calE_{11} \Lambda_o\| ^2 =O_p(\frac 1 {\No} ), \quad
\frac 1 {\Tm } \|\frac 1 {\No } \calE_{21} \Lambda_o\| ^2 =O_p(\frac 1 {\No} ),\quad
 \frac 1 {\To} \|\frac 1 {\Nm } \calE_{12} \Lambda_m\| ^2 =O_p(\frac 1 {\Nm} ).
\end{equation}
It follows that
\[ \frac 1 {T} \|\frac 1 {N} \calE_{11}\Lambda_o\| ^2 =\frac{\To}T (\frac {\No } N)^2 O_p(1/\No )=O_p(1/N) \]
and similarly
\[ \frac 1 {T} \|\frac 1 {N} \calE_{21}\Lambda_o\| ^2
=O_p(1/N), \text{ and }
 \frac 1 {T} \|\frac 1 {N} \calE_{12}\Lambda_m\| ^2
 =O_p(1/N). \]

Next consider
\[ \frac 1 T \|\frac 1 N u \Lambda_m \| ^2 =\frac 1 T (\frac {\Nm } N)^2 \|\frac 1 {\No }
\calE_{21} \Lambda_o B (\Lambda_m'\Lambda_m/\Nm )\| ^2 \le \frac {\Tm } T (\frac {\Nm } N)^2
\Big[ \frac 1 {\Tm } \|\frac 1 {\No }
\calE_{21} \Lambda_o\| ^2 \Big] \|B (\Lambda_m'\Lambda_m/\Nm )\|. \]
Using $B (\Lambda_m'\Lambda_m/\Nm )=O_p(1)$ (an $r$ by $r$ matrix), and (\ref{1stBlock}),
\[ \frac 1 T \|\frac 1 N u \Lambda_m \| ^2=\frac {\Tm } T (\frac {\Nm } N)^2 \frac 1 {\No } O_p(1)=O_p(1/\No ) \]
(which can be much smaller than $O_p(1/\No )$ , depending on $\Tm /T$ and $\Nm /N$).
Consider
\[ \frac 1 T \|\frac 1 N v \Lambda_m \| ^2  =
  \frac 1 T \| \frac 1 N  \frac 1 {\To} F_m^0 A F_o^{0\prime}\calE_{12} \Lambda_m \| ^2 .  \]
The $r \times r$ matrix satisfies $\frac 1 N  \frac 1 {\To}  F_o^{0\prime} \calE_{12}\Lambda_m =O_p((N\To)^{-1/2})$. Thus
 \[ \frac 1 T \|\frac 1 N v \Lambda_m \| ^2\le (\frac 1 T \|F_m^0A\| ^2)\|\frac 1 N  \frac 1 {\To}  F_o^{0\prime}\calE_{12} \Lambda_m \| ^2=\frac 1 {N\To} O_p(1), \]
 this term is dominated by others.
Summarizing results, we have
\[ \frac 1 T \| \frac 1 N e^\dag \Lambda\| ^2= O_p(1/N) + \frac {\Tm } T (\frac {\Nm } N)^2 \frac 1 {\No } O_p(1)=O_p(1/\No ). \]
Now
\[ \frac 1 T \|F\Lambda'e^{\dag\prime} \tildeF^+/(NT)\|^2 \le (\frac 1 T \|F\|^2)(\frac 1 T \|\tildeF^+\|^2)
\Big(\frac 1 T \| \frac 1 N \Lambda' e^{\dag\prime} \|^2 \Big)=O_p(1/\No ), \]
\[ \frac 1 T \|e^\dag \Lambda F'\tildeF^+/(NT)\|^2\le \Big(\frac 1 T \|\frac 1 N e^\dag \Lambda\|^2\Big) \|F'\tildeF^+/T\|^2 =O_p(1/\No )O_p(1) =O_p(1/\No ). \]

 Bai and Ng (2002) show that,
 \begin{equation}\label{eeF}  \frac 1 T \| e e' \tilde F/(NT) \| ^2 = O_p(1/T+1/N). \end{equation}
 Here we will show
\begin{equation}\label{eedagF} \frac 1 T \| e^\dag  e^{\dag\prime} \tildeF^+ /(NT) \| ^2 =O_p(1/\No +1/\To).  \end{equation}
Notice
\[   e^\dag e^{\dag \prime} \tildeF^+ =
\begin{bmatrix} (\calE_{11}\calE_{11}'+\calE_{12}\calE_{12}')\tildeF^+_o & [\calE_{11}\calE_{21}'+\calE_{12}(u+v)']\tildeF^+_m\\
[\calE_{21}\calE_{11}'+(u+v)\calE_{12}']\tildeF^+_o   & [\calE_{22}\calE_{22}'+(u+v)(u+v)']\tildeF^+_m \end{bmatrix}. \]
Consider the first block. Let $e_o=(\calE_{11}, \calE_{12}) $ (a matrix of dimension $\To\times N$), then
the first block is $e_o e_o' \tildeF^+_o$, which is a subblock of $ee'\tildeF^+$. Thus, from (\ref{eeF}),
\[  \frac 1 T \|e_o e_o'\tildeF^+_o/(NT)\| ^2\le \frac 1 T \|ee'\tildeF^+/(NT)\| ^2 =O_p(1/T+1/N) \]
[in fact, $(\To/T) O_p(1/T+1/N)$].
Next consider the off-diagonal block. Noticing
$\|\calE_{11}\calE_{21}'\tildeF^+_m \| ^2 \le \|\calE_{11}\calE_{21}'\| ^2 \|\tildeF^+_m\| ^2$, where
$\calE_{11}$ is $\To\times \No $, and $\calE_{21}'$ is $\No \times \Tm $, and
$\|\calE_{11}\calE_{21}'\|^2 =\sum_{t=1}^{\To}\sum_{h=1}^{\Tm } \Big(\sum_{j=1}^{\No } e_{jt}e_{j,\To+h}\Big)^2$. Thus
\[ \frac 1 T  \|\calE_{11}\calE_{21}'\tildeF^+_m /(NT)\|^2\le \frac {\To}T \frac {\Tm } T (\frac {\No } N)^2
\Big( \frac 1 {\To \Tm } \sum_{t=1}^{\To}\sum_{h=1}^{\Tm } \Big( \frac 1 {\No }\sum_{j=1}^{\No } e_{jt}e_{j,\To+h}\Big)^2 (\|\tildeF^+_m\|^2/T) \]
\[ =\frac {\To}T \frac {\Tm } T (\frac {\No } N) O_p(1/N). \]
Here for simplicity, we assume the non-overlapping errors are uncorrelated.
The block $\frac 1 T \| \|\calE_{21}\calE_{11}'\tildeF^+_o /(NT)\|^2$ is of the same order of magnitude as above.
Next,
\[ \|\calE_{22}\calE_{22}'/(\Tm \Nm )\|^2  =O_p(1/\Tm +1/\Nm ), \]
this implies
\[ \frac 1 T \|\calE_{22}\calE_{22}'\tildeF^+_m/(TN)\|^2  =\frac {\Tm }T \frac {\Nm } N O_p(1/T+1/N). \]
Next consider
\[ \calE_{12}u'\tildeF^+_m =\frac 1 {\No } \calE_{12}\Lambda_m B \Lambda_o'\calE_{21}' \tildeF^+_m. \]
Thus
\[ \frac 1 T \|\calE_{12}u'\tildeF^+_m/(NT)\|^2\le \frac 1 {T^2}(\frac {\Nm } N)^2\|\frac 1 {\Nm }\calE_{12}\Lambda_m \|^2
(\|B\|^2) \|\frac 1 {\No } \Lambda_o'\calE_{21}'\|^2 (\|\tildeF^+_m\|^2/T) \]
\[ =(\frac {\Nm } N)^2 (\frac {\To} T) (\frac {\Tm } T) O_p(1/(\No \Nm )) =(\frac {\Nm } N) (\frac {\To} T) (\frac {\Tm } T)  \frac 1 N O_p(1/\No ), \]
the last equality uses results (\ref{1stBlock}).
Similarly,  $\frac 1 T \|\calE_{12}v'\tildeF^+_m/(NT)\|^2$ is negligible.

Next consider $\frac 1 T \| (u+v)(u+v)' \tildeF^+_m/(NT)\| ^2$.
The dominating terms are $\frac 1 T \|uu'\tildeF^+_m/(NT)\| ^2$ and $\frac 1 T \|vv'\tildeF^+_m/(NT)\| ^2$.
We analyze each of them. Note
$ uu'\tildeF^+_m = \frac 1 {\No ^2} \calE_{21}\Lambda_o B (\Lambda_m' \Lambda_m) B \Lambda_o'\calE_{21}'\tildeF^+_m$, and
 $B(\Lambda_m'\Lambda_m/\Nm )B=O_p(1)$, thus
\[  \frac 1 T \|uu'\tildeF^+_m/(NT)\| ^2 \le (\frac {\Nm } N)^2 (\frac {\Tm }T)^2 \Big(\frac 1 {\Tm }
\|\frac 1 \No  \calE_{21} \Lambda_o \| ^2\Big)^2 (\frac 1 T \|\tildeF^+_m\| ^2) O_p(1)=O_p(1/\No ^2) \]
where we use (\ref{1stBlock}) and  $\frac 1 T \|\tildeF^+_m\| ^2 =O_p(1)$.
Next
\[ vv'\tildeF^+_m/(NT) = \frac 1 {\To^2} F_m^0 A (F_o^{0\prime} \calE_{12} \calE_{12}'F_o^0) A (F_m^{0\prime} \tildeF^+_m/T)/N. \]
The three $r\times r$ matrices satisfy
\[ \frac 1 {\Nm } \frac 1 {\To} (F_o^{0\prime} \calE_{12} \calE_{12}'F_o^0)=\frac 1 {\Nm } \sum_{i=\No +1}^N
\Big[\frac 1 {\To} \Big(\sum_{t=1}^{\To} F_t^0 e_{it}\Big)\Big(\sum_{t=1}^{\To} F_t^0 e_{it}\Big)'\Big]=O_p(1), \]
 $(F_m^{0\prime}\tildeF^+_m/T)=O_p(1)$, and $\|A\|=O_p(1)$. Thus
 \[  \frac 1 T \|vv'\tildeF^+_m/(NT)\| ^2=(\frac {\Nm } N)^2 \frac{\Tm } T \frac 1 {\To^2} (\frac 1 {\Tm } \|F_m^0\| ^2) O_p(1) =(\frac {\Nm } N)^2 ( \frac{\Tm } T) O_p(1/\To^2). \]
 Summarizing results gives us (\ref{eedagF}).
This completes the proof of Lemma \ref{lem:averageRate}(i).

\paragraph{Proof of Lemma \ref{lem:averageRate}(ii).}  Left multiplying (\ref{identity1}) by $F'$ on each side and dividing by $T$,
\[ (F'F/T)(\Lambda'\Lambda/N)(F'\tildeF^+/T)+(F'F/T)\Lambda'e^{\dag\prime} \tildeF^+/(NT) +F'e^\dag \Lambda F'\tildeF^+/(NT^2)
\]
\[ +F' e^\dag e^{\dag\prime}\tildeF^+/(NT^2) =(F'\tildeF^+/T) {\tilde D_r}^2  .\]
By the argument of  Bai (2003), it is sufficient to prove each of  the last 3 terms on the left hand side converges in probability to zero. That is,
\begin{equation} \label{Q1}\Lambda'e^{\dag\prime} \tildeF^+/(NT) \pconv  0 ,\end{equation}
\begin{equation}\label{Q2} F'e^\dag \Lambda /(NT) \pconv 0 ,\end{equation}
\begin{equation} \label{Q3}  F' e^\dag e^{\dag\prime}\tildeF^+/(NT^2) \pconv 0 .\end{equation}
Consider (\ref{Q2}) first.
\[ F'e^\dag \Lambda/(NT) = [ F_o'\calE_{11}\Lambda_o +F_m'\calE_{21}\Lambda_o+ F_o'\calE_{12}\Lambda_m +F_m'(u+v)\Lambda_m]/(NT). \]
The first 3 terms are each $O_p((NT)^{-1/2})$. The last term is
\[ F_m'u\Lambda_m/(NT) + F_m'v \Lambda_m/(NT) =a+b, \text{ with} \]
\[ a= \frac 1 {\No } F_m' \calE_{21} \Lambda_o B \Lambda_m' \Lambda_m /(NT) =\frac {\Nm } N \frac 1 {T\No } F_m' \calE_{21} \Lambda_o
 B (\Lambda_m'\Lambda_m/\Nm )  =\frac {\Nm } N O_p(( T\No )^{-1/2}), \]
  \[ b= \frac 1 {\To} F_m^{0\prime} F_m^0 A F_o^{0\prime} \calE_{12}\Lambda_m /(NT)
  = \frac {\Tm } T ( F_m^{0\prime} F_m^0/\Tm ) A  \frac 1 {N\To} F_o^{0\prime} \calE_{12} \Lambda_m  =\frac {\Tm } T O_p((N\To)^{-1/2}). \]
  This proves (\ref{Q2}). For (\ref{Q1}),
\[ \Lambda' e^{\dag\prime} \tildeF^+=\Lambda' e^{\dag\prime} (\tildeF^+-F H^++FH^+). \]
  After dividing by $NT$, the term involving $\tildeF^+-F H^+$ is negligible, using Lemma \ref{lem:averageRate}(i).
  Notice $\Lambda' e^{\dag\prime}  F H^+ /(NT)$
  is equal to the transpose of (\ref{Q2}) (ignoring $H$), this proves (\ref{Q1}).
  Similarly,
  \[ F^{\prime} e^\dag e^{\dag\prime}\tildeF^+/(NT^2)= F' e^\dag e^{\dag\prime}(\tildeF^+-FH^++FH^+)/(NT^2). \]
  The term involving $(\tildeF^+-FH^+)$ is negligible. It suffices to show
  $F^{\prime} e^\dag e^{\dag\prime} F/(NT^2) \pconv 0$.
  Bai and Ng (2002) proved $F'ee'F/(NT^2)$ to be $o_p(1)$. Given the difference between $e$ and $e^\dag$,
 it remains to show
 \[ F_m'(u+v)(u+v)' F_m/(NT^2) =F_m'(uu'+uv'+vu'+vv')F_m)/(NT^2) =o_p(1). \]
 The dominating term $(F_m'vv'F_m)/(NT^2)$ is
 \[ F_m'vv'F_m/(NT^2)= \frac 1 {\To^2} \Fop_mF_m^0 A F_o^{0\prime} \calE_{12} \calE_{12}' F_o^0 A \Fop_mF_m^0  =(\frac {\Tm }  T)^2 \frac {\Nm } N  \frac 1 {\To} O_p(1) =o_p(1).  \]
 Summarizing results we have proved
 \[ (F'F/T)(\Lambda'\Lambda/N)(F'\tildeF^+/T)+ o_p(1) =(F'\tildeF^+/T) {\tilde D_r^2}  . \]
The remaining proof is similar to the proof of Proposition 1 of Bai (2003). This implies that $F'\tildeF^+/T$ converges to $\mathbbQ_r$, and ${\tilde D_r}$ converges to $\mathbbD_r$.
This complete the proof of Lemma \ref{lem:averageRate}.

\begin{proposition} (asymptotic representation for $\tildeF^+$). Under Assumptions of A-B
 \label{hatX-thm1}
 \begin{itemize}
\item[(a)] for $t\le \To$,
$ \tildeF^+_t-H^{+\prime} F_t^0= {\tilde D_r}^{-2} (\hFplusp F^0/T) \frac 1 N \sum_{k=1}^N \lambdaUpperCase_k^0 e_{kt} +\hat \xi_{NT,t}$;
\item[(b)] for $t>\To$,
$\tildeF^+_t-H^{+\prime} F_t^0  ={\tilde D_r}^{-2} (\hFplusp F^0/T) {\mathbf B_\Lambda} \frac 1 {\No } \sum_{i=1}^{\No } \lambdaUpperCase_i^0 e_{it} + \hat \xi_{NT,t}+ O_p((N\To)^{-1/2})$,
where $\hat \xi_{NT,t}=O_p(1/\min\{\No , \To\})$ uniformly in $t$, and the $r\times r$ matrix ${\mathbf B_\Lambda}$ is defined as
\[ {\mathbf B_\Lambda}= \frac {\No }  N I_r +\frac {\Nm} N \Big(  \frac{\Lambda_m^{0\prime}\Lambda_m^0 }{\Nm} \Big)  \Big( \frac{\Lambda_o'\Lambda_o}{\No} \Big)^{-1}. \]
 \end{itemize}
\end{proposition}
\noindent
{\bf Proof of Proposition \ref{hatX-thm1}}.  Let $e_{it}^\dag$ denote the $(i,t)$th entry of $e^\dag$.
\[ \tildeF^+_t -H^{+\prime}F_t^0= {\tilde D_r}^{-2}\frac 1 {NT} \sum_{s=1}^T \sum_{i=1}^N \tildeF^+_s\lambdaUpperCase_i'e_{is}^\dag F_t^0
+ {\tilde D_r}^{-2} (\hFplusp F/T) \frac 1 N \sum_{i=1}^N \lambdaUpperCase_i e_{it}^\dag +{\tilde D_r}^{-2} \frac 1 {NT} \sum_{i=1}^N\sum_{s=1}^T \tildeF^+_s e_{is}^\dag e_{it}^\dag. \]
We can show that the first and the last terms on the right hand side are
$O_p(1/\delta_{\No ,\To}^2)$, the limiting distribution is determined by the second term. That is,
\begin{equation} \label{Fplus-rep}   \tildeF^+_t -H^{+\prime}F_t^0=
{\tilde D_r}^{-2} (\hFplusp F/T) \frac 1 N \sum_{i=1}^N \lambdaUpperCase_i e_{it}^\dag +O_p(1/\delta_{\No ,\To}^2). \end{equation}
For $t\le \To$, then $e_{it}^\dag =e_{it}$ for all $i$. This gives part (a) of Proposition \ref{hatX-thm1}.
But for $t>\To$
\[ e_{it}^\dag = \left\{ \begin{array}{ll} e_{it} & i\le \No \\
 u_{it}+v_{it} & i>\No   \end{array} \right.  \]
 Plugging in  $e_{it}^\dag$ into the preceding formula we obtain
\[\tildeF^+_t-H^{+\prime} F_t^0= {\tilde D_r}^{-2} (\hFplusp F^0/T) \frac 1 N \Big(\sum_{i=1}^{\No } \lambdaUpperCase_i^0 e_{it}
+\sum_{i=\No +1}^N \lambdaUpperCase_i^0 (u_{it}+v_{it}) \Big)+\hat \xi_{NT,t}.\] But
\begin{eqnarray*}
   \frac 1 N \sum_{i=\No +1}^N \lambdaUpperCase_i^0 (u_{it}+v_{it})& =&\frac 1 N \sum_{i=\No +1}^N \lambdaUpperCase_i^0 \lambdaUpperCase_i^{0\prime}
 (\Lambda_o'\Lambda_o/\No )^{-1}\frac 1 {\No } \sum_{k=1}^{\No } \lambdaUpperCase_k^0 e_{kt} \\
&& +\frac 1 N  \sum_{i=\No +1}^N \lambdaUpperCase_i^0 F_t^{0\prime} (F_o^{0\prime} F_o^0/\To)^{-1}\frac 1 {\To} \sum_{s=1}^{\To} F_s^0 e_{is} =I_1 +I_2.
\end{eqnarray*}
 The first term on the right is equal to  (note $\Nm=N-\No$)
 \[ I_1= \frac {\Nm } N \Big(  \frac{\Lambda_m^{0\prime}\Lambda_m^0 }{\Nm} \Big)  \Big( \frac{\Lambda_o'\Lambda_o}{\No} \Big)^{-1}\frac 1 {\No } \sum_{k=1}^{\No } \lambdaUpperCase_k^0 e_{kt}, \]
 where $\frac{\Lambda_m^{0\prime}\Lambda_m^0 }{\Nm} =\frac 1 {N-\No } \sum_{i=\No +1}^N \lambdaUpperCase_i^0 \lambdaUpperCase_i^{0\prime}$ and
term $I_2$ is negligible because it can be rewritten as
\[ I2= F_t^{0\prime} (F_o^{0\prime} F_o^0/\To)^{-1} \frac 1 {N \To}\sum_{i=\No +1}^N \sum_{s=1}^{\To} F_s^0 \lambdaUpperCase_i^0 e_{is} =O_p( (N\To)^{-1/2}).\]
Note $F_t^{0\prime} (F_o^{0\prime} F_o^0/\To)^{-1}F_s^0$ is a scalar and is commutable with $\lambdaUpperCase_i^0$. Summarizing result, for $t>\To$,
\begin{equation} \label{non-stationary-Lambda}
 \tildeF^+_t-H^{+\prime} F_t^0= {\tilde D_r}^{-2} (\hFplusp F^0/T) \Big[ \frac {\No }  N I_r +\frac {\Nm} N \Big(  \frac{\Lambda_m^{0\prime}\Lambda_m^0 }{\Nm} \Big)  \Big( \frac{\Lambda_o'\Lambda_o}{\No} \Big)^{-1} \Big]  \frac 1 {\No }  \sum_{i=1}^{\No } \lambdaUpperCase_i^0 e_{it} + \hat \xi_{NT,t}+ O_p((N\To)^{-1/2}).
\end{equation}
This gives the representation in part (b).
Under block stationarity assumption, $\mathbf B_\Lambda \pconv \mathbb B_\Lambda=I_r$,
then the representation is simplified to
\begin{equation}
\tildeF^+_t-H^{+\prime} F_t^0={\tilde D_r}^{-2} (\hFplusp F^0/T) \frac 1 {\No } \sum_{i=1}^{\No } \lambdaUpperCase_i^0 e_{it} + \hat \xi_{NT,t}+ O_p((N\To)^{-1/2}).
\end{equation}
It is important to note that the limiting distribution of
$\tildeF^+_t-H^{+\prime} F_t^0$ is determined by $\frac 1 {\No} \sum_{i=1}^{\No } \lambdaUpperCase_i^0 e_{it}$, and
the convergence rate is $\sqrt{\No }$.

\begin{corollary} \label{hatX-coro1} Under Assumptions A-C, and $\mathbf B_\Lambda\pconv \mathbb B_\Lambda$, then\\
(a) for $t\le \To$,
$ \sqrt{N}(\tildeF^+_t -H^{+\prime}F_t^0) \dconv N(0, \mathbbD_r^{-2} \mathbbQ_r \Gamma_t \mathbbQ_r' \mathbbD_r^{-2})$,  \\
(b) for $t>\To$,
$\sqrt{\No }(\tildeF^+_t -H^{+\prime}F_t^0) \dconv N(0, \mathbbD_r^{-2} \mathbbQ_r \mathbb B_\Lambda \Gamma_{ot}  \mathbb B_\Lambda' \mathbbQ_r' \mathbbD_r^{-2})$.
\end{corollary}

\paragraph{Proof of Corollary \ref{hatX-coro1}.}
Using  ${\tilde D_r}\rightarrow^p \mathbbD_r$ and  $\hFplusp F/T\pconv \mathbb Q_r$, part (a) follows from $\frac 1 {\sqrt{N} } \sum_{i=1}^N \lambdaUpperCase_i e_{it} \dconv N(0,\Gamma_t)$ by Assumption A.
Part (b) follows from $\frac 1 {\sqrt{\No } } \sum_{i=1}^{\No } \lambdaUpperCase_i e_{it} \dconv N(0,\Gamma_{ot})$ by Assumption C.

\begin{proposition} \label{hatX-thm2} (asymptotic representation of $\hat \Lambda_i^+$) Under Assumptions A-B,
\begin{itemize}
\item[(a)] for $i\le \No, $
$ \hL_i-G^+ \lambdaUpperCase_i^0= H^{+\prime}  \frac 1 T \sum_{t=1}^T F_t^0 e_{it} +\hat \eta_{NT,i}  $;
\item[(b)] for $i>\No $,
$  \hL_i-G^+ \lambdaUpperCase_i^0= H^{+\prime}  \frac 1 T \Big( \sum_{t=1}^{\To} F_t^0 e_{it} +\sum_{t=\To+1}^T F_t^0 (u_{it}+v_{it})\Big)+ \hat \eta_{NT,i}
 =H^{+\prime}  \mathbf B_F \frac 1 {\To} \Big( \sum_{t=1}^{\To} F_t^0 e_{it} \Big) +\hat \eta_{NT,i} +O_p((T \No )^{-1/2}), $
where $\hat \eta_{NT,i}=O_p(1/\No +1/\To)$ uniformly in $i$, and the $r\times r$ matrix $\mathbf B_F$ is defined as
\[ \mathbf B_F =\frac{\To} T I_r+ \frac {\Tm} T \Big(\frac 1 {\Tm} \sum_{t=\To+1}^T F_t^0F_t^{0\prime}\Big) (F_o^{0\prime} F_o^0/\To)^{-1}. \]

\end{itemize}
\end{proposition}

\paragraph{Proof of Proposition \ref{hatX-thm2}.} Given Proposition \ref{hatX-thm1}, the proof of Proposition \ref{hatX-thm2} invokes some symmetry arguments, as in Bai (2003). The details are omitted. Part (b) of the Proposition uses
\begin{eqnarray*}
   \frac 1 T \sum_{t=\To+1}^T F_t^0 u_{it}& =&O_p((T\No )^{-1/2}), \\
 \frac 1 T \sum_{t=\To+1}^T F_t^0 v_{it}&=& \frac {T-\To} T \Big(\frac 1 {T-\To} \sum_{t=\To+1}^T F_t^0F_t^{0\prime}\Big) (F_o^{0\prime} F_o^0/\To)^{-1}\frac 1 {\To} \sum_{s=1}^{\To} F_s^0 e_{is}.
\end{eqnarray*}
 This implies
$ \hL_i- G^+ \lambdaUpperCase_i^0= H^{+\prime} \mathbf B_F \frac 1 {\To}  \sum_{t=1}^{\To} F_t^0 e_{it} +\hat \eta_{NT,i}+O_p((T\No )^{-1/2})$ for part (b).
\begin{corollary} \label{hatX-coro2} Under Assumptions A-C, and $\mathbf B_F\pconv \mathbb B_F$, then\\
(a) for $i\le \No $,
$ \sqrt{T} (\hL_i-G^+ \lambdaUpperCase_i^0) \dconv N(0, \mathbbQ_r^{\prime -1} \Phi_i \mathbbQ_r^{-1})$\\
(b) for $i >  \No $,
$ \sqrt{\To} (\hL_i-G^+ \lambdaUpperCase_i^0) \dconv  N(0, \mathbbQ_r^{\prime -1} \mathbb B_F  \Phi_{oi} \mathbb B_F' \mathbbQ_r^{-1})$.
\end{corollary}

\paragraph{Proof of Corollary \ref{hatX-coro2}.}  This follows from the asymptotic representation in Proposition \ref{hatX-thm2}.



\paragraph{Proof of Proposition \ref{prop:hatX-coro1}.}   Parts (a) and (b) of the proposition are implied by Corollary \ref{hatX-coro1}
and parts (c) and (d) of the proposition are implied by Corollary \ref{hatX-coro2}.

\paragraph{Proof of Proposition \ref{thm:missC}.} We shall prove the proposition without Assumption D.  We assume $\mathbf B_\Lambda
\pconv \mathbb B_\Lambda$ and $\mathbf B_F\pconv \mathbb B_F$ (see, equation (\ref{BLBF-limit}), the limit may not be an identity matrix).
Consider the case for $i\le \No $ and $t\le \To$. We need to show
\begin{equation} \label{thm:missC-eqn1} (\frac 1 N V_{it}+ \frac 1 T W_{it})^{-1/2} (\imputedCit-C_{it}^0) \dconv N(0,1),
\end{equation}
where $V_{it}=\lambdaUpperCase^{\prime}_i \Sigma_\Lambda^{-1}\Gamma_t \Sigma_\Lambda^{-1}\lambdaUpperCase_i$
and $W_{it}= F'_t(\Sigma_F^{-1} \Phi_i \Sigma_F^{-1}) F_t$.

To see this, rewrite
 the representations in part (a) of Proposition \ref{hatX-thm1} as
\[ G^{+\prime} \tildeF^+_t-F_t^0 = (\Lambda'\Lambda/N)^{-1} \frac 1 N
\sum_{k=1}^N \lambdaUpperCase_k e_{kt} + O_p(1/\delta_{\No ,\To}^2), \]
where again $G^+=(H^+)^{-1}$. This follows from
\[ G^{+\prime}  {\tilde D_r}^{-2}(\hFplusp  F^0/T)=G^{+\prime}  {\tilde D_r}^{-2}(\hFplusp  F^0/T)(\Lambda'\Lambda/N)(\Lambda'\Lambda/N)^{-1} =G^{+\prime} H^{+\prime} (\Lambda'\Lambda/N)^{-1}=
(\Lambda'\Lambda/N)^{-1}. \]
Similarly rewrite the representation in part (a) of Proposition \ref{hatX-thm2} as
\[ H^+\tLp_i -\lambdaUpperCase_i^0=(F'F/T)^{-1} \frac 1 T \sum_{s=1}^T F_s e_{is} +O_p(1/\delta_{\No ,\To}^2). \]
Here we have used $H^+H^{+\prime}=(F'F/T)^{-1} + O_p(1/\delta_{\No ,\To}^2)$.
Thus
\begin{eqnarray} \imputedCit-C_{it}&=& \tildeF_t^{+\prime}\tLp_i-F_t'\lambdaUpperCase_i
=\tildeF_t^{+\prime}G^{+} H^+\tLp_i-F_t'\lambdaUpperCase_i
 =(\tildeF_t^{+\prime} G^{+} -F_t' +F_t') (H^+\tLp_i-\lambdaUpperCase_i+\lambdaUpperCase_i)-C_{it} \nonumber \\ \label{cit-cit-a}
 &=&(\tildeF_t^{+\prime} G^{+} -F_t')(H^+\tLp_i-\lambdaUpperCase_i)
  +F_t'(F'F/T)^{-1} \frac 1 T \sum_{s=1}^T F_s e_{is}
  \\&&+\lambdaUpperCase_i' (\Lambda'\Lambda/N)^{-1} \frac 1 N \sum_{k=1}^N \lambdaUpperCase_k e_{kt}
  +O_p(1/\delta_{\No ,\To}^2)= I1+I2+I3+I4,\nonumber
\end{eqnarray}
 where both $I1$ and $I4$ are $O_p(1/\delta_{\No ,\To}^2)$. By assumption,
 $\sqrt{N}O_p(1/\delta_{\No ,\To}^2)\rightarrow 0$, and $\sqrt{T}O_p(1/\delta_{\No ,\To}^2)\rightarrow 0$,
 so $I1$ and $I4$ are dominated terms.
  Also by assumption, $N^{-1/2} \sum_{k=1}^N \lambdaUpperCase_k e_{kt}\dconv N(0,\Gamma_t)$,  it follows that, conditional on $\lambdaUpperCase_i$ (if it is random),
 \[ \sqrt{N} \, \lambdaUpperCase_i' (\Lambda'\Lambda/N)^{-1}  \frac 1 N \sum_{k=1}^N \lambdaUpperCase_k e_{kt} \dconv N(0, V_{it}),\]
 and similarly,
 \[ \sqrt{T} \, F_t'(F'F/T)^{-1} \frac 1 T \sum_{s=1}^T F_s e_{is}  \dconv N(0,W_{it}).\]
 The two limiting distributions are asymptotically independent, this implies (\ref{thm:missC-eqn1}) (see the proof of Theorem 3 in Bai, 2003).

Now consider $t> \To$, still with $i\le \No$. From part (b) of Proposition \ref{hatX-thm1}, rewrite  the representation as
\[  G^{+\prime} \tildeF^+_t-F_t^0 = (\Lambda'\Lambda/N)^{-1} \mathbf B_\Lambda \frac 1 {\No }
\sum_{k=1}^{\No } \lambdaUpperCase_k e_{kt} + O_p(1/\delta_{\No ,\To}^2). \]
Using the same argument as in the proof of (\ref{thm:missC-eqn1}), we have
\[ \imputedCit-C_{it} = F_t'(F'F/T)^{-1} \frac 1 T \sum_{s=1}^T F_s e_{is}
   +  \lambdaUpperCase_i' (\Lambda'\Lambda/N)^{-1} \mathbf B_\Lambda \frac 1 {\No } \sum_{k=1}^{\No } \lambdaUpperCase_k e_{kt} +O_p(1/\delta_{\No ,\To}^2). \]
The limit of the first term was considered earlier. The second term, multiplying $\sqrt{\No }$, is asymptotically normal $N(0, V_{it}^o)$, where
\[ V_{it}^0= \Lambda_i' \Sigma_\Lambda^{-1} \mathbb B_\Lambda \Gamma_{ot} \mathbb B_\Lambda'\Sigma_\Lambda^{-1}\Lambda_i. \]
The two terms are asymptotically independent. Thus
$ (\frac 1 T W_{it} +\frac 1 {\No} V_{it}^o)(\imputedCit-C_{it})\dconv N(0,1).$

 The proof for the block $i>\No, t\le \To$ is the same. The asymptotic representation becomes
 \[ \imputedCit-C_{it} = F_t'(F'F/T)^{-1} \mathbf B_F \frac 1 {\To} \sum_{s=1}^{\To} F_s e_{is}
   +  \lambdaUpperCase_i' (\Lambda'\Lambda/N)^{-1} \frac 1 N \sum_{k=1}^N \lambdaUpperCase_k e_{kt} +O_p(1/\delta_{\No ,\To}^2). \]
This implies $(\frac 1 {\To} W_{it}^o+ \frac 1 {N} V_{it})(\imputedCit-C_{it})\dconv N(0,1)$, where
$W_{it}^o=F_t'\Sigma_F^{-1}\mathbb B_F \Phi_{oi} \mathbb F_F'\Sigma_F^{-1} F_t$.
Finally, for $i>\No$ and $t>\To$ (the missing block), the asymptotic representation is
 \begin{equation}\label{Cit+-Cit+} \imputedCit-C_{it} =  F_t'(F'F/T)^{-1} \mathbf B_F \frac 1 {T_o} \sum_{s=1}^{T_o} F_s e_{is}
 +\lambdaUpperCase_i' (\Lambda'\Lambda/N)^{-1} \mathbf B_\Lambda \frac 1 {\No } \sum_{k=1}^{\No } \lambdaUpperCase_k e_{kt} +O_p(1/\delta_{\No ,\To}^2).\end{equation}
 This implies  $  (\frac 1 {\To} W_{it}^o+\frac 1 {\No} V_{it}^o)^{-1/2} (\imputedCit-C_{it}^0) \dconv N(0,1)$.

\paragraph{Proof of Corollary \ref{cor:avgC}.}
Consider the block defined by $i\le N_o, t\le T_o$.
From the representation in (\ref{cit-cit-a}). Term I2 is
$O_p(T^{-1/2})$ for each $i$ and $t$. Taking squares and then averaging over this block gives the rate $O_p(1/T)$. The square root of the average is
$O_p(T^{-1/2})$. Similarly, averaging the squares of term I3 gives $O_p(1/N)$. The square root of this average is $O_p(N^{-1/2})$.
Term I1 and term I4 are both uniformly bounded by $O_p(1/\delta_{\No ,\To}^2)$. Thus its Frobenius norm over the corresponding blocks is still of this magnitude. This implies
$ \frac {\|\tilde C_1^+-C_{1}^0\|} {\sqrt{N_o T_o}} =O_p(\frac 1{\sqrt{N}})+O_p(\frac 1{ \sqrt{T}})+ O_p(\delta_{N_o,T_o}^{-2})$.
The proofs for other blocks are the same. For example, for the block $i>\No$ and $t>\To$, we use
(\ref{Cit+-Cit+}) to obtain
$ \frac {\|\tilde C_4^+-C_4^0\|} {\sqrt{N_m T_m}} =O_p(\frac 1{\sqrt{N_o}})+O_p(\frac 1{ \sqrt{T_o}})+ O_p(\delta_{N_o,T_o}^{-2}) $.
Corollary 1 is obtained by averaging the four blocks,  the weight for each block corresponds to the block size.

\end{document}